\documentclass[preprint,12pt,authoryear]{elsarticle}

\usepackage{amsmath}
\usepackage{amssymb}
\usepackage{graphicx}
\usepackage{booktabs}
\usepackage{multirow}
\usepackage{array}
\usepackage{url}
\usepackage{subcaption}
\usepackage{threeparttable}
\usepackage{float}
\usepackage{tabularx}
\usepackage[a4paper,top=2.5cm,bottom=2.5cm,left=2.2cm,right=2.2cm]{geometry}

\usepackage[colorlinks=true,
            citecolor=blue,
            linkcolor=blue,
            urlcolor=blue]{hyperref}

\usepackage{newunicodechar}
\newunicodechar{≥}{\geq}

\newcommand{\degC}{\ensuremath{^\circ}C}   
\newcommand{\ACE}{\mathrm{ACE}}            

\usepackage[section]{placeins}   

\journal{Tropical Cyclone Research and Review}

\begin{document}

\begin{frontmatter}


\title{ENSO-linked redistribution of post-monsoon tropical cyclone energy between the Arabian Sea and the Bay of Bengal: 1982--2025}


 \author[inst1]{Prashant Saini}
 \author[inst1]{Rishikesh Yadav\corref{cor1}}
 \cortext[cor1]{Corresponding author}
 \ead{rishikesh@iitmandi.ac.in}
 \author[inst1]{Sarita Azad}


 \affiliation[inst1]{
     organization={School of Mathematical and Statistical Sciences,
     Indian Institute of Technology Mandi},
     city={Mandi},
     state={Himachal Pradesh},
     country={India}
 }


\begin{abstract}
Tropical cyclones in the North Indian Ocean (NIO) form in two sub-basins, the Arabian Sea (AS) and the Bay of Bengal (BoB), which are usually studied separately, with their relationship to the El Ni\~no-Southern Oscillation (ENSO) treated as stationary. We develop a joint inter-basin statistical framework and apply it to 416 disturbances from quality-controlled IBTrACS best tracks for 1982--2025, with harmonised IMD and JTWC intensities, ERA5 and ERSSTv5. The framework tests whether ENSO changes the basin total or redistributes energy between the sub-basins, separates its effect on formation from that on intensification, attributes each sub-basin's response to environmental terms along individual tracks, and tests stationarity and the record length needed to detect trends. In October-December, ENSO redistributes rather than changes the total: for a one-standard-deviation increase in Ni\~no-3.4 (about 1 $^\circ$C), accumulated cyclone energy (ACE) is multiplied by 1.89 in the AS and by 0.63 in the BoB, while the NIO total shows no detectable change (factor 0.99 [0.74, 1.32]). ENSO acts on intensification, with no detectable effect on formation: the odds that a cyclonic storm becomes very severe rise about fourfold in the AS and halve in the BoB, through lower vertical wind shear over the AS and weaker low-level vorticity over the BoB. The relationship has changed: El Ni\~no lowered total NIO energy in 1982--2003, but in 2004--2025 AS gains balanced BoB losses (change p = 0.016). Very severe storms and rapid intensification increased in the AS only, and no 15-year window detects this increase, so about three decades of data are needed. The post-monsoon IOD signal can hardly be separated from ENSO, and August Ni\~no-3.4 gives modest skill for the AS-BoB partition but none for the basin total.
\end{abstract}


\end{frontmatter}


\section{Introduction}
\label{sec:intro}

Tropical cyclones (TCs) in the North Indian Ocean (NIO) are few in number but highly destructive. The basin contributes only a small fraction of global TC activity but its storms impact some of the most densely populated and low-lying coastlines on Earth, from Bangladesh, Myanmar and eastern India around the Bay of Bengal (BoB) to western India, Pakistan, Oman and Yemen around the Arabian Sea (AS). Storm surge and floods from the 1970 Bhola cyclone, Nargis in 2008 and Amphan in 2020 indicate the potential for a single NIO storm to become a regional calamity, and the cost of such events has risen with coastal vulnerability
\citep{mohapatra2014tropical,tiwari2022appraisal,singh2022review,wu2024escalation}. The NIO has two main cyclone seasons. Most cyclones form during the pre-monsoon (April–June) and post-monsoon (October–December) seasons. During the southwest monsoon (June--September), strong upper-level easterlies
produce large vertical wind shear (VWS), which makes conditions unfavourable for cyclone formation and intensification
\citep{evan2011climatology,li2013bimodal,wang2022climate}.
The two sub-basins also differ strongly. The BoB produces two to three times as many disturbances and cyclonic storms (CS) as the AS, which is cooler, drier in the mid-troposphere and, until recently, much less active \citep{evan2011climatology,ng2012interannual,liu2021understanding,pathirana2022tropical}. Understanding what controls the balance between the two sub-basins is therefore as important for risk assessment as understanding the activity of the basin as a whole.

The large-scale controls on TC genesis and intensification are well established: warm sea-surface temperature (SST) and upper-ocean heat content, a moist mid-troposphere, cyclonic low-level vorticity and weak VWS \citep{gray1968global,demaria1994statistical,demaria1996effect}. Genesis potential indices combine these ingredients and allow their contributions to be separated \citep{camargo2007use,suneeta2025role}. In the NIO, VWS sets
the seasonal windows for intensification \citep{wang2022climate}, while SST and the subsurface ocean modulate storm intensity and life cycle, especially
in the BoB \citep{balaguru2014increase,sebastian2015impact,mohanty2019role}. The same ingredients are perturbed by modes of interannual climate variability, which is the basis for seasonal TC outlooks in the region \citep{biswas2018principal,jaiswal2024seasonal}.

The El Ni\~no--Southern Oscillation (ENSO) is the most studied of these modes. In the BoB, post-monsoon TC frequency, accumulated cyclone energy (ACE) and power dissipation are lower in El Ni\~no than in La Ni\~na seasons, and genesis shifts with the phase
\citep{girishkumar2012influences,ng2012interannual,mahala2015impacts,bhardwaj2019tropical}. \citet{roose2022enso} showed that this ENSO effect is confined to low-latitude genesis and is linked to equatorial wind and vorticity anomalies, and
\citet{sumesh2013tropical} reported different responses during El Ni\~no Modoki years. In the AS, El Ni\~no and positive Indian Ocean Dipole (IOD) seasons have been associated with stronger and longer-lived post-monsoon storms \citep{yuan2013north,almahri2025teleconnections}, and AS-active and BoB-active seasons have been contrasted in terms of the northeast monsoon and an IOD-like moisture redistribution \citep{cheung2019comparison}. Few studies have examined both sub-basins together. \citet{das2023tropical} found opposite ENSO responses of AS and BoB ACE over 29 years and higher total NIO ACE in La Ni\~na years, which suggests that ENSO modulates the basin total. To our knowledge, no study has tested formally whether ENSO changes the basin total or only its partition between the two sub-basins. At the global scale, ENSO has only a weak influence on rapid intensification (RI) in the Indian Ocean \citep{silvers2025global}.

The role of the IOD is less clear. The IOD is the leading mode of interannual SST variability in the tropical Indian Ocean \citep{saji1999dipole}, and positive events have been linked to enhanced AS and suppressed BoB activity \citep{yuan2013north,mahala2015impacts}. However, the IOD and ENSO are strongly coupled in boreal autumn, when many positive IOD events develop together with El Ni\~no \citep{saji1999dipole,cai2019pantropical,Dutta2024_ENSO_IOD}.
Studies that composite TC activity by combined ENSO--IOD phase therefore rest on few seasons per category and cannot separate the two modes. Whether the IOD has any influence on NIO TCs that is independent of ENSO, and in which
season, has not been established.

On longer time scales, TC activity in the NIO has changed. Extremely severe cyclonic storms have become more frequent over the AS \citep{murakami2017increasing,swapna2022increasing}, an increase that has been attributed to anthropogenic warming, to aerosol-induced weakening of the monsoon circulation and shear \citep{evan2011arabian}, and to rapid Indian Ocean warming \citep{roxy2015drying,roxy2020indian}. \citet{deshpande2021changing} reported a 52\% increase in AS CS and a 150\% increase in very severe cyclonic storms (VSCS) between 1982--2000 and 2001--2019, and the AS genesis potential has increased \citep{vidya2023intensification}. Post-monsoon BoB storms have also intensified \citep{balaguru2014increase}, and RI events have become more common in offshore regions worldwide \citep{Kossin2020_PNAS,li2023recent}. In the AS, the number of RI events has roughly doubled since 2013, together with warmer SST and stronger low-level vorticity \citep{li2025sharp}. Yet trend estimates for the NIO differ in sign and magnitude between studies \citep{tiwari2022understanding,tiwari2022assessing,thomas2022recent,abhinav2025warming}. These studies use different periods, epochs and best-track sources. Most compare two fixed epochs or use records of 15--40 years, and the sensitivity of the trend to the length and placement of the record has not been quantified. NIO best tracks also mix intensities from the India Meteorological Department (IMD, 3-min sustained wind) and the Joint Typhoon Warning Center (JTWC, 1-min), whose averaging periods are not equivalent
\citep{knapp2010international,harper2010guidelines}. If the two sources are combined without harmonisation, part of an apparent trend can arise from changes in which agency provides the intensity.

Finally, the relationship between ENSO and the Indian monsoon system is not stationary. The ENSO--Indian summer monsoon rainfall relationship has varied over the past century and may shift under warming \citep{azad2016possible}, and ENSO's influence on the East Asian summer monsoon has weakened since the early 2000s \citep{yu2025weakened}. If the ENSO--TC relationship in the NIO has changed in the same way, results from studies covering different periods would not be expected to agree. In the BoB, the ENSO effect on post-monsoon storms is known to be stronger in warm phases of the Pacific Decadal Oscillation \citep{girishkumar2015influence}, but whether the relationship has changed in recent decades, and how the two
sub-basins share in any such change, has not been examined.

These studies leave five questions open. First, because the AS and the BoB have almost always been analysed separately, it is not known whether ENSO changes the \emph{total} post-monsoon cyclone energy of the NIO or only moves it between the two sub-basins, and whether that partition can be anticipated before the season begins. Second, seasonal counts and ACE combine genesis, intensification and storm duration, so it is not known at which stage of the life cycle ENSO acts. Third, basin-mean composites do not show which environmental ingredient each sub-basin responds to, or what individual storms experience along their tracks. Fourth, the autumn coupling of ENSO and the IOD leaves their separate roles, and any ENSO-independent IOD influence, ambiguous. Fifth, the relationship is usually treated as stationary, the dependence of
trend detection on record length has not been quantified for the NIO, and genesis-density maps
\citep{wahiduzzaman2017climatological,wahiduzzaman2020kernel} have not been tested for field significance \citep{wilks2016stippling}.

We address these questions with 44 years (1982--2025) of quality-controlled International Best Track Archive for Climate Stewardship (IBTrACS) best tracks over a basin-wide domain, with IMD and JTWC intensities
harmonised, combined with ERA5 reanalysis and ERSSTv5 SST
\citep{knapp2010international,hersbach2020era5,huang2017extended}. Our central hypothesis is that ENSO \emph{redistributes} post-monsoon TC energy between the AS and the BoB rather than modulating the NIO total. To test it we build a joint inter-basin statistical framework. Compared with existing approaches,
the study is new in six respects.
\begin{enumerate}
\renewcommand{\labelenumi}{(\roman{enumi})}
\item \emph{A joint inter-basin test of modulation against redistribution.}
Earlier studies have related ENSO to TC activity in each sub-basin separately, or compared the two with composites. We model the two together and use two complementary indices, the AS--BoB difference in ACE and the NIO total. Modulation predicts a response of the total, redistribution a response
of the difference only, so the choice becomes an explicit statistical test, which we check with eleven robustness tests.
\item \emph{Formation separated from intensification.} Earlier work relates ENSO to seasonal counts and ACE, which mix genesis, intensification and duration. We model the probability of each step of the storm life cycle (depression to CS, CS to VSCS, and rapid intensification) with storm-level logistic regression, and split seasonal ACE exactly into frequency, duration and intensity terms.
\item \emph{Storm-level attribution of the environmental pathway.} Earlier work compares basin-mean environments between ENSO phases, usually for one sub-basin. We separate the genesis-potential terms for both sub-basins in one
framework, add the environment sampled along each storm's track, and estimate how much of the ENSO effect on intensification that environment carries.
\item \emph{ENSO separated from the IOD, season by season.} Earlier work composites TC activity by combined ENSO--IOD phase, with few seasons in each class. We use an ENSO-independent DMI, partial correlations and an ENSO$\times$IOD interaction model, separately for the pre- and post-monsoon
seasons.
\item \emph{A formal test of non-stationarity and of record length.} Earlier work compares fixed epochs and, apart from decadal modulation linked to the Pacific Decadal Oscillation in the BoB, treats the ENSO relationship as stationary. We test whether the relationship changed between the two halves
of the record, check that the result does not depend on the split year, and quantify how trend detection depends on the length and start year of the record, with single-agency checks and field-significance tests for spatial patterns.
\item \emph{Prediction of the inter-basin partition.} Existing seasonal outlooks target basin-wide activity. We assess, with leave-one-out cross-validation and skill scores judged against the ``always BoB-dominated'' reference, how far the AS--BoB partition can be predicted from August Ni\~no-3.4.
\end{enumerate}

The remainder of the paper is organised as follows. Section~\ref{sec:study_area} describes the study area and the analysis domain. Section~\ref{sec:data_methods} presents the data, preprocessing and statistical methods. Section~\ref{sec:result} presents the results, Section~\ref{sec:discussion} discusses them in relation to previous work and their limitations, and Section~\ref{sec:conclusion} summarises the main conclusions.

\section{Study area}
\label{sec:study_area}

\begin{figure}[t]
\centering
\includegraphics[width=0.95\textwidth]{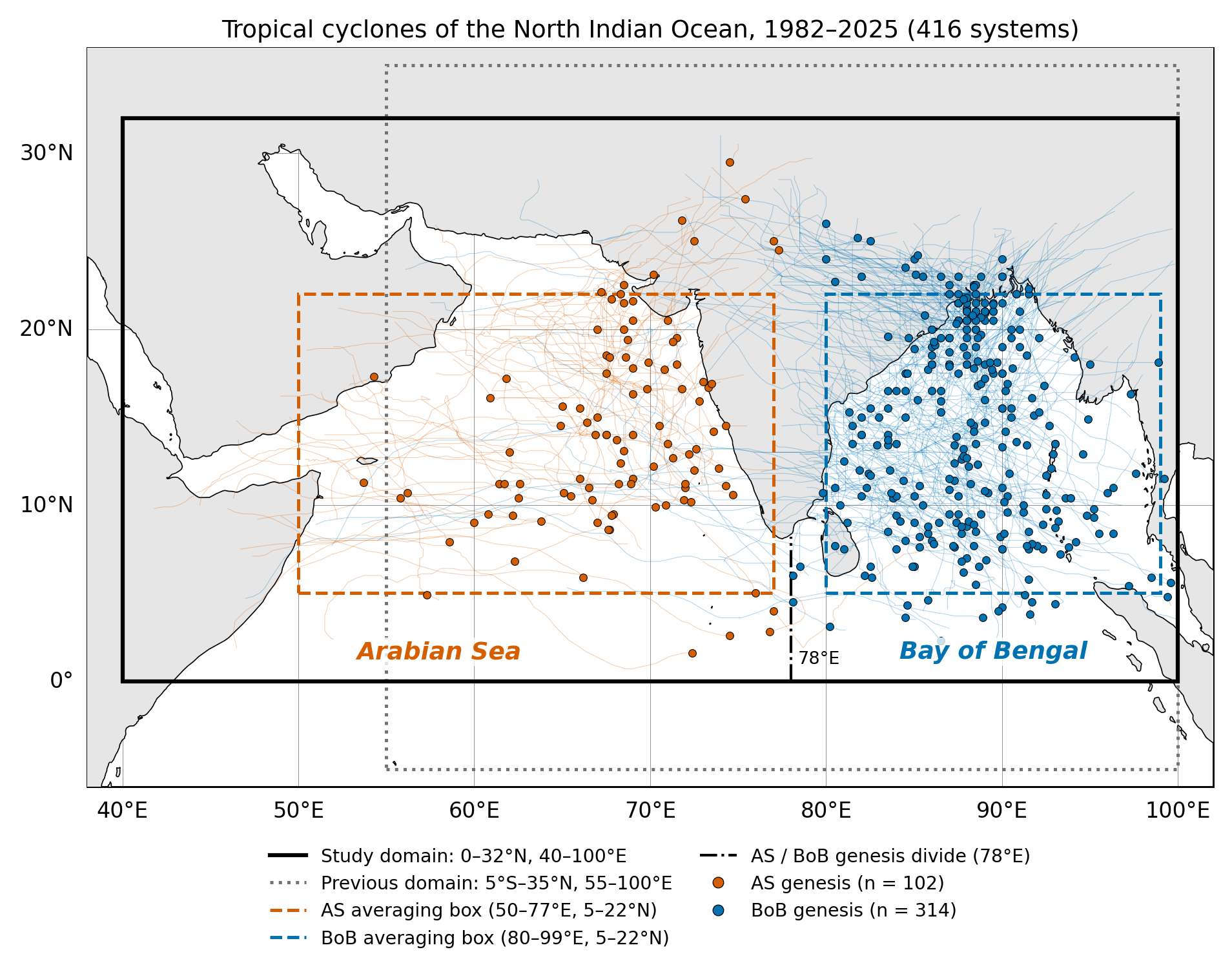}
\caption{Study domain (solid box, 0--32$^\circ$N, 40--100$^\circ$E) with the tracks (thin lines) and genesis positions (dots; first fix with wind $\geq$17\,kt) of the 416 systems analysed for 1982--2025, coloured by the sub-basin in which they formed (AS, $n=102$; BoB, $n=314$). The dash-dotted
line marks the AS/BoB boundary at 78$^\circ$E south of India, and the dashed boxes are the regions used for basin-mean environmental diagnostics. The dotted box marks a narrower 5$^\circ$S--35$^\circ$N, 55--100$^\circ$E domain, shown for comparison, which leaves out the western AS.}
\label{fig:study_area}
\end{figure}

The NIO lies between the Horn of Africa and the Malay
Peninsula, and the Indian subcontinent splits it into two sub-basins: the AS to the west and the BoB, together with the
Andaman Sea, to the east (Fig.~\ref{fig:study_area}). Land closes the basin to the north. Cyclones that form here therefore either decay over the sea or make landfall on the coasts of India, Sri Lanka, Bangladesh and Myanmar around the BoB, or of India, Pakistan, Oman, Yemen and Somalia around the AS.

We work in a domain of 0--32$^\circ$N and 40--100$^\circ$E, which contains every NIO track fix in IBTrACS for 1982--2025 (0.7--31.0$^\circ$N, 41.8--100.0$^\circ$E). Placing the western edge at 40$^\circ$E keeps the western AS, the Gulf of Aden and the Somali coast in the analysis. This part of the basin is small but not empty: 518 fixes and the lifetime-maximum-intensity (LMI) positions of 11 systems lie west of
55$^\circ$E, and Cyclone Sagar (May 2018) formed at 53.7$^\circ$E. A domain starting at 55$^\circ$E would leave out this western part of the AS (dotted box in Fig.~\ref{fig:study_area}). The southern edge follows the equator, because no NIO system has been recorded in the Southern Hemisphere, and the northern edge at 32$^\circ$N lies just beyond the northernmost fix, which is over land.

Sub-basins follow the IBTrACS convention, with the AS/BoB boundary at 78$^\circ$E south of the peninsula. A storm is assigned to the sub-basin in which it forms. ACE, on the other hand, is assigned to the sub-basin in which each 6-hourly fix lies, so a BoB storm that crosses southern India and re-emerges over the AS adds to AS energy from that point on. Of the 416 systems analysed, 314 formed in the BoB and
102 in the AS. For basin-mean environmental quantities we use two boxes that cover the main regions of genesis and intensification, 50--77$^\circ$E, 5--22$^\circ$N for the AS and 80--99$^\circ$E, 5--22$^\circ$N for the BoB, with land points masked.

The two sub-basins have different climates. The BoB is warmer, its middle troposphere is moister, and the monsoon trough over its northern part feeds it a steady supply of low-pressure systems. The AS is cooler and drier, and its low-level vorticity is weak except in the transition seasons
(Appendix Fig.~\ref{fig:app_env}; \citealp{evan2011climatology,liu2021understanding,pathirana2022tropical}). In both, cyclone activity is tied to the monsoon: the high shear of the southwest monsoon (June--September) confines cyclones mainly to the pre-monsoon (April--June, AMJ) and post-monsoon (October--December, OND)
seasons \citep{li2013bimodal}. Over 1982--2025, 57\% of cyclonic storms formed in OND and 32\% in AMJ (Appendix Fig.~\ref{fig:app_clim}a). The AS shows the clearest double peak, in May--June and October--November, with very little in between.

\section{Data and methods}
\label{sec:data_methods}

\subsection{Physical background}
\label{subsec:physical_background}

The variables analysed here follow from the physics of TC development. Warm SST supplies, through evaporation, the moisture whose condensation releases the latent heat that drives convection, weak VWS lets the vortex stay upright, a moist middle troposphere limits the drying of convective cells by entrainment, and cyclonic low-level vorticity gives
a disturbance its initial spin \citep{gray1968global,holton2013introduction}. We therefore represent the four ingredients by SST (ERSSTv5), 200–850 hPa VWS, 700 hPa relative humidity and 850 hPa relative vorticity (ERA5), which are also the terms of the genesis potential index used in Section~\ref{subsec:statistical_analysis}. Once a storm moves over land it loses its energy source and decays. ENSO and the IOD alter these ingredients through the Walker
circulation, equatorial winds and the distribution of SST and moisture across the basin \citep{saji1999dipole,camargo2007use,roose2022enso}. There is no a priori reason for the response to have the same sign in the AS and the BoB, and this is the starting point of the redistribution hypothesis tested here.

\subsection{Data}
\label{subsec:data_sources}

\begin{table}[htbp]
\centering
\caption{Datasets used in this study.}
\label{tab:data_sources}
\small
\begin{tabularx}{\textwidth}{@{}>{\raggedright\arraybackslash}p{4.2cm} >{\raggedright\arraybackslash}p{3.6cm} l >{\raggedright\arraybackslash}X@{}}
\toprule
\textbf{Dataset} & \textbf{Variable} & \textbf{Period} & \textbf{Use} \\
\midrule
IBTrACS v04r01 (NI) & tracks, IMD and JTWC winds & 1982--2025 & TC metrics \\
NOAA PSL Ni\~no-3.4 (HadISST) & SST anomaly & 1982--2025 & ENSO \\
NOAA PSL DMI (HadISST1.1) & SST gradient & 1982--2025 & IOD \\
NOAA ERSSTv5 & monthly SST, 2$^\circ$ & 1982--2025 & SST composites, trends, genesis SST \\
ERA5 (C3S) & $u$, $v$ at 200 and 850\,hPa & 1982--2025 & VWS, low-level wind \\
ERA5 (C3S) & RH at 700\,hPa & 1982--2025$^a$ & mid-level moisture \\
ERA5 (C3S) & relative vorticity at 850\,hPa & 1982--2025$^a$ & low-level spin \\
\bottomrule
\end{tabularx}

\vspace{2pt}
{\footnotesize $^a$Assembled from two ERA5 retrievals on the same grid
(RH: 1985 and 2025 retrieved separately; vorticity: 2025 retrieved separately).}
\end{table}

\textit{Tropical cyclones.} Tracks come from the International Best Track Archive for Climate Stewardship (IBTrACS v04r01, NI basin) \citep{knapp2010international,IBTrACS_NCEI_2024} for 1982--2025 (Table~\ref{tab:data_sources}). We start in 1982, at the beginning of the satellite era, when NIO cyclone intensities have been estimated routinely from geostationary and polar-orbiting imagery and are more homogeneous than
in earlier decades \citep{evan2011climatology,Kossin2020_PNAS}. Even so, IMD
intensities are missing for 15--25\% of storms in the 1980s and 1990s, and an earlier start would increase the reliance on converted JTWC winds. ERA5 is also consistent in its use of satellite data from 1979 onwards \citep{hersbach2020era5}, and 1982 is the usual start year of satellite-era NIO studies \citep{deshpande2021changing,almahri2025teleconnections}, which
makes our results directly comparable with others. The record ends with the 2025 season. Seven 2025 systems were still marked as provisional when the data were retrieved; we keep them and test their influence separately (Table~\ref{tab:robust}). For each fix, IBTrACS gives two intensity estimates: the 3-min sustained wind of the IMD, and the 1-min sustained wind of the JTWC.

\textit{Climate indices.} ENSO is represented by the monthly Ni\~no-3.4 SST anomaly and the IOD by the Dipole Mode Index (DMI) \citep{saji1999dipole}, both computed from HadISST \citep{rayner2003global} and distributed by NOAA PSL \citep{NOAA_PSL_SST}. Seasonal indices are three-month means. We use September--November (SON) for the OND cyclone season and March--May (MAM) for AMJ, because the leading season does not overlap with the cyclone response and is the season a forecaster would have. El Ni\~no and La Ni\~na seasons are those with SON Ni\~no-3.4 $\geq$+0.5\,\degC\ and $\leq-$0.5\,\degC\ (14 and 16 seasons), and positive and negative IOD seasons those with SON DMI beyond $\pm$0.4\,\degC, roughly one standard deviation.

\textit{SST.} Monthly SST for 1982--2025 is taken from NOAA ERSSTv5 \citep{huang2017extended} on its 2$^\circ$ grid, which gives a homogeneous record for the full period. Anomalies are departures from the 1982--2025 monthly climatology.

\textit{Atmospheric environment.} We use monthly ERA5 fields
\citep{hersbach2020era5,era5_pressure_monthly} at 0.25$^\circ$: zonal and meridional wind at 200 and 850\,hPa, relative humidity at 700\,hPa and relative vorticity at 850\,hPa, all for 1982--2025. For relative humidity and vorticity, two downloads of the same ERA5 product were merged; no month is missing in any field. VWS is computed from the 200 and 850 hPa winds as,
\begin{equation}
\mathrm{VWS}=\left[(u_{200}-u_{850})^2+(v_{200}-v_{850})^2\right]^{1/2}.
\end{equation}

All fields were averaged in 4$\times$4 blocks onto a common 1$^\circ$ grid over 0–30°N, 40–100°E, which covers both analysis boxes and all fixes over water, and VWS was computed from these monthly-mean 1$^\circ$ winds.

\subsection{Data preprocessing and quality control}
\label{subsec:preprocessing}

\begin{table}[t]
\centering
\caption{Quality-control and preprocessing steps applied to the
IBTrACS NI record (1982--2025).}
\label{tab:qc}
\small
\begin{tabularx}{\textwidth}{@{}>{\raggedright\arraybackslash}X >{\raggedleft\arraybackslash}p{4.3cm}@{}}
\toprule
\textbf{Step} & \textbf{Effect} \\
\midrule
Raw record (NI basin, 1982--2025) & 15\,102 fixes, 462 systems \\
Remove \texttt{spur-merge} track segments & $-$66 fixes ($-$1 system) \\
Duplicate SID--time records & none found \\
Restrict to the study domain 0--32$^\circ$N, 40--100$^\circ$E & all fixes retained \\
Flag provisional 2025 tracks (kept; tested separately) & 7 systems \\
West Pacific entrants (genesis east of 100$^\circ$E): kept for ACE, excluded from genesis statistics & 43 systems \\
Systems with no wind estimate at any fix (15 of them also West Pacific entrants) & 17 systems \\
Intensity: IMD 3-min wind; otherwise JTWC 1-min wind converted with Eq.~(\ref{eq:conv}) & 62.9\% IMD, 31.4\% converted \\
Energy metrics on synoptic times only (00, 06, 12, 18 UTC) & 7\,426 off-synoptic fixes excluded \\
Categories from wind thresholds (Table~\ref{tab:imd_categories}), not IMD grade labels & consistent across 1982--2025 \\
\midrule
Analysed systems: 461 $-$ 43 $-$ 17 + 15 & 416 (AS 102, BoB 314) \\
\quad of which $\geq$CS / $\geq$SCS / $\geq$VSCS / RI & 195 / 104 / 64 / 58 \\
\bottomrule
\end{tabularx}
\end{table}

Table~\ref{tab:qc} lists the preprocessing steps. Spur tracks, which duplicate parts of main tracks, were removed, and all fixes lie inside the study domain. Storms that formed east of 100$^\circ$E and entered the NIO from the West Pacific (43 systems) were kept for the energy budget, since
the energy they produce over the BoB is real, but were left out of every statistic that depends on where a storm formed. 17 systems had no wind value at any fix; 15 of these are also West Pacific entrants, mostly weak remnants that crossed Myanmar or Thailand. Excluding these 45 systems from the 461 in the record leaves 416 systems for the storm-level statistics; the seasonal ACE series also include the 28 West Pacific entrants that have wind values. Two further choices matter for the homogeneity of the record.

\textit{Intensity harmonisation.} IMD winds are missing for about 37\% of fixes (51\% in 1982–2003 and 25\% in 2004–2025). We filled these gaps with JTWC winds converted to the IMD scale by a linear fit to the 3\,119 synoptic fixes that have both estimates and an IMD wind of at least 17\,kt (Appendix Fig.~\ref{fig:app_clim}c):
\begin{equation}
v_{\mathrm{IMD}} = 3.86 + 0.812\,v_{\mathrm{JTWC}}, \qquad r = 0.90.
\label{eq:conv}
\end{equation}
The fitted slope implies a larger reduction for strong storms than a change of averaging period alone would give \citep{harper2010guidelines}, so the fit also absorbs systematic differences between the two agencies' estimates. For this reason we use the empirical relation rather than a fixed WMO factor. Remaining gaps of up to two consecutive fixes (6 h) in the merged wind were filled by linear interpolation. Because converted winds make up a larger share early in the record, the main results were recomputed with IMD-only and with JTWC-only winds
(Table~\ref{tab:robust}).

\textit{Categories.} IMD introduced the ESCS grade only in 2015, so the grade labels in the archive are not consistent over 1982--2025. Each system was instead given the category of its lifetime maximum wind (Table~\ref{tab:imd_categories}).

Genesis is defined as the first fix inside the NIO with a wind of at least 17\,kt. The LMI is the highest merged wind, and its position is the LMI location. RI is an increase
of at least 30\,kt in 24\,h between synoptic fixes \citep{kaplan2003large}.

\begin{table}[htbp]
\centering
\caption{IMD classification of cyclonic disturbances, used to assign categories from lifetime maximum wind.}
\label{tab:imd_categories}
\small
\begin{tabularx}{\textwidth}{@{}Xcc@{}}
\toprule
\textbf{Category} & \textbf{Wind (kt)} & \textbf{Wind (km h$^{-1}$)} \\
\midrule
Depression (D) & 17--27 & 31--49 \\
Deep Depression (DD) & 28--33 & 50--61 \\
Cyclonic Storm (CS) & 34--47 & 62--88 \\
Severe Cyclonic Storm (SCS) & 48--63 & 89--117 \\
Very Severe Cyclonic Storm (VSCS) & 64--89 & 118--166 \\
Extremely Severe Cyclonic Storm (ESCS) & 90--119 & 167--221 \\
Super Cyclonic Storm (SuCS) & $\geq$120 & $\geq$222 \\
\bottomrule
\end{tabularx}

\vspace{2pt}
{\footnotesize Source: \citet{imd_faq_tropical_cyclones_2023}.}
\end{table}

\subsection{Metrics}

Accumulated cyclone energy \citep{bell2000climate} is
\begin{equation}
\ACE = 10^{-4}\sum v_{\max}^2 \quad (10^{4} \mathrm{kt}^2), \qquad v_{\max}\geq 34\ \mathrm{kt},
\end{equation}
summed over synoptic fixes and assigned to the sub-basin in which each fix lies. Because ACE is computed from the merged IMD-scale winds, its values are lower than ACE based on 1-min winds. The central quantity of the study is the difference in OND energy between the two sub-basins,
\begin{equation}
\Delta \ACE_y = \ACE^{\mathrm{AS}}_{y} - \ACE^{\mathrm{BoB}}_{y},
\label{eq:dace}
\end{equation}
which we examine alongside the total, $\ACE^{\mathrm{AS}}_{y}+\ACE^{\mathrm{BoB}}_{y}$. The two hypotheses make different predictions. If ENSO \emph{modulated} basin-wide activity, it would change the total; if it \emph{redistributed} activity, it would change $\Delta \ACE$ and would change $\Delta \ACE$ and leave the total largely unchanged, which we judge from the confidence interval of the total effect rather than from its $p$-value alone. We use a difference rather than a ratio because 15 of the 44 OND seasons have no AS ACE at all.

\subsection{Statistical methods}
\label{subsec:statistical_analysis}

\textit{Trends.} Annual counts are modelled with negative-binomial generalised linear models (GLMs) with a log link and a linear time term, with the dispersion parameter estimated by maximum likelihood. We report rate ratios per decade with Wald 95\% confidence intervals. For ACE and SST we use the Mann--Kendall test with the \citet{hamed1998modified} correction for autocorrelation, and Sen's slope for the magnitude \citep{sen1968estimates}.

\textit{Dependence on record length.} To see how much a trend depends on the period analysed, we computed it for every window of 10--44 years and every start year within 1982--2025 (Fig.~\ref{fig:window}). Windows of ten years are too short to estimate overdispersion or autocorrelation reliably, so for
this analysis counts were fitted with a Poisson GLM and ACE was tested with the original Mann--Kendall test. Both are, if anything, more likely than the full models to detect a trend, so a failure to detect one in short windows is not caused by this choice.

\textit{ENSO and IOD relationships.} Associations are measured with Spearman rank correlation, which is robust to the skewed distribution and the few extreme seasons of the ACE series; the many zero values enter as tied ranks. Two-sided $p$-values come from 10,000 permutations and 95\% intervals from 5,000 bootstrap resamples. As a check on dependence between consecutive years, intervals were also computed with a moving-block bootstrap with 3-year blocks \citep{kunsch1989jackknife}; Table~\ref{tab:robust}. Effect sizes are estimated with
quasi-Poisson GLMs of seasonal ACE on standardised indices, in which the Pearson dispersion inflates the standard errors. Because ENSO and the IOD are correlated in autumn, we also define an ENSO-independent DMI, $\mathrm{DMI_{ind}}$, as the residual of the DMI regressed on Ni\~no-3.4, fitted separately for each calendar month. We further report partial rank correlations and a GLM that includes an ENSO$\times$IOD interaction. The redistribution hypothesis rests on two pre-specified tests, the association of Niño-3.4 with $\Delta\ACE$ and with the total; the remaining tests describe how the redistribution arises.

\textit{Formation versus intensification.} For each OND storm, logistic regressions on standardised SON Ni\~no-3.4 model four conditional probabilities: P($\geq$CS\,$|$\,D), P($\geq$SCS\,$|$\,CS), P($\geq$VSCS\,$|$\,CS) and P(RI\,$|$\,CS). An ENSO effect on the first points to control of formation, and effects on the others to control of
intensification. We also split seasonal ACE in each sub-basin into three factors,
\begin{equation}
ACE = 4\times10^{-4}\,N\,D\,\overline{v^2} \;\propto\; N\,D\,\overline{v^2},
\label{eq:ace_decomp}
\end{equation}
where $N$ is the number of storms reaching 34\,kt in the sub-basin per season, $D$ the number of storm-days per storm (6-hourly fixes of at least 34\,kt divided by four) and $\overline{v^2}$ the mean squared wind (kt$^2$) over those fixes. Because ACE is proportional to the product of the three factors, the constant cancels when two groups of seasons are compared, and the El Ni\~no--La Ni\~na difference in $\ln \ACE$ is the exact sum of a frequency, a duration and an intensity term,
\begin{equation}
\Delta\ln \ACE = \Delta\ln N + \Delta\ln D + \Delta\ln\overline{v^2}.
\end{equation}
Uncertainty in each term is estimated by resampling El Ni\~no and La Ni\~na seasons with replacement (3\,000 bootstrap samples).

\textit{Spatial patterns and field significance.} El Ni\~no--La Ni\~na differences in TC-days and in intensifying fixes (a gain of at least 5\,kt in 6\,h) were computed on a 2.5$^\circ$ grid. Pointwise $p$-values come from 5,000 permutations of the El Niño and La Niña labels among the 30 ENSO seasons, and because many grid points are tested at once we control the Benjamini--Hochberg false-discovery rate at $q=0.10$ \citep{benjamini1995controlling,wilks2016stippling}. Genesis and LMI densities were estimated with a Gaussian kernel (bandwidth factor 0.25), following \citet{wahiduzzaman2017climatological}, and scaled to storms per season. Differences between 1982--2003 and 2004--2025 were tested by permuting whole years between the two epochs
10,000 times, again with FDR control over active ocean points.

\textit{Environmental diagnostics.} OND El Ni\~no--La Ni\~na composites of the environmental fields were tested at each grid point with two-sample $t$-tests, with FDR control. Basin means over the AS and BoB boxes, with land masked and grid points weighted by the cosine of latitude, were compared
between phases with Mann--Whitney tests and related to Ni\~no-3.4 with Spearman correlations. To identify which ingredient each sub-basin responds to, we followed \citet{camargo2007use} and evaluated separately the three
atmospheric terms of the logarithm of the Emanuel--Nolan genesis potential index: the vorticity term $\tfrac{3}{2}\ln|10^5\eta_{850}|$, where $\eta_{850}$ is the absolute vorticity at 850\,hPa, the humidity term $3\ln(\mathrm{RH}_{700}/50)$ and the shear term $-2\ln(1+0.1\,\mathrm{VWS})$. On this logarithmic scale a change of 0.1 in any term corresponds to a change of about 10\% in the index. The potential-intensity term is not included, because computing it requires full vertical profiles of temperature and humidity, which were not retrieved for this study; its main driver, SST, is analysed directly instead.

Basin means describe the background state but not necessarily the conditions that individual storms meet. We therefore also sampled the environment along the track of each OND CS. The sampled segment runs from the first fix at 34\,kt to the time of LMI (at least 24\,h), and only fixes over water are used. At each synoptic fix we took the mean of the 3$\times$3 one-degree grid points centred on the storm in the monthly field for that month, and then averaged these values along the segment, giving one value of VWS, RH$_{700}$ and $\zeta_{850}$ per storm. The SST anomaly at genesis was obtained in the same way from ERSSTv5, using the 3$\times$3 grid boxes around the genesis position in the month of genesis. To estimate how much of the ENSO effect on P($\geq$VSCS\,$|$\,CS) is carried by the along-track environment, we fitted two logistic models, one with Ni\~no-3.4 alone ($\beta_0$) and one that also includes along-track VWS and RH$_{700}$ ($\beta_1$). The share explained is $1-\beta_1/\beta_0$, and its 95\% interval comes from 2,000 bootstrap resamples of the storms. Because coefficients of logistic models are not collapsible, adding predictors of the outcome can enlarge the ENSO coefficient even when they carry none of its effect, so this share is a conservative, approximate measure. Resamples in which a model did not converge were discarded.

\textit{Stationarity and predictability.} The stability of the ENSO relationship is examined with running Spearman correlations in 21-year windows. Differences between the two 22-year halves (1982--2003 and 2004--2025) are tested by permutation, using 5,000 random splits of the 44 years into two groups, and by Fisher's $z$. Predictability is assessed with a leave-one-out cross-validated linear hindcast of OND $\Delta \ACE$ and of total ACE from August Ni\~no-3.4, which is available about a month before the season begins. Skill is measured by the cross-validated correlation, the mean-squared skill score (MSSS) relative to a cross-validated climatology, and the Heidke skill score (HSS), with an event defined as a season in the upper quartile of the observed $\Delta\ACE$ and forecast when the hindcast is in the upper quartile of the hindcasts.

\section{Results}
\label{sec:result}

\subsection{Changes in NIO TC activity, 1982--2025}
\label{subsec:trends}

\begin{figure}[t]
\centering
\includegraphics[width=0.92\textwidth]{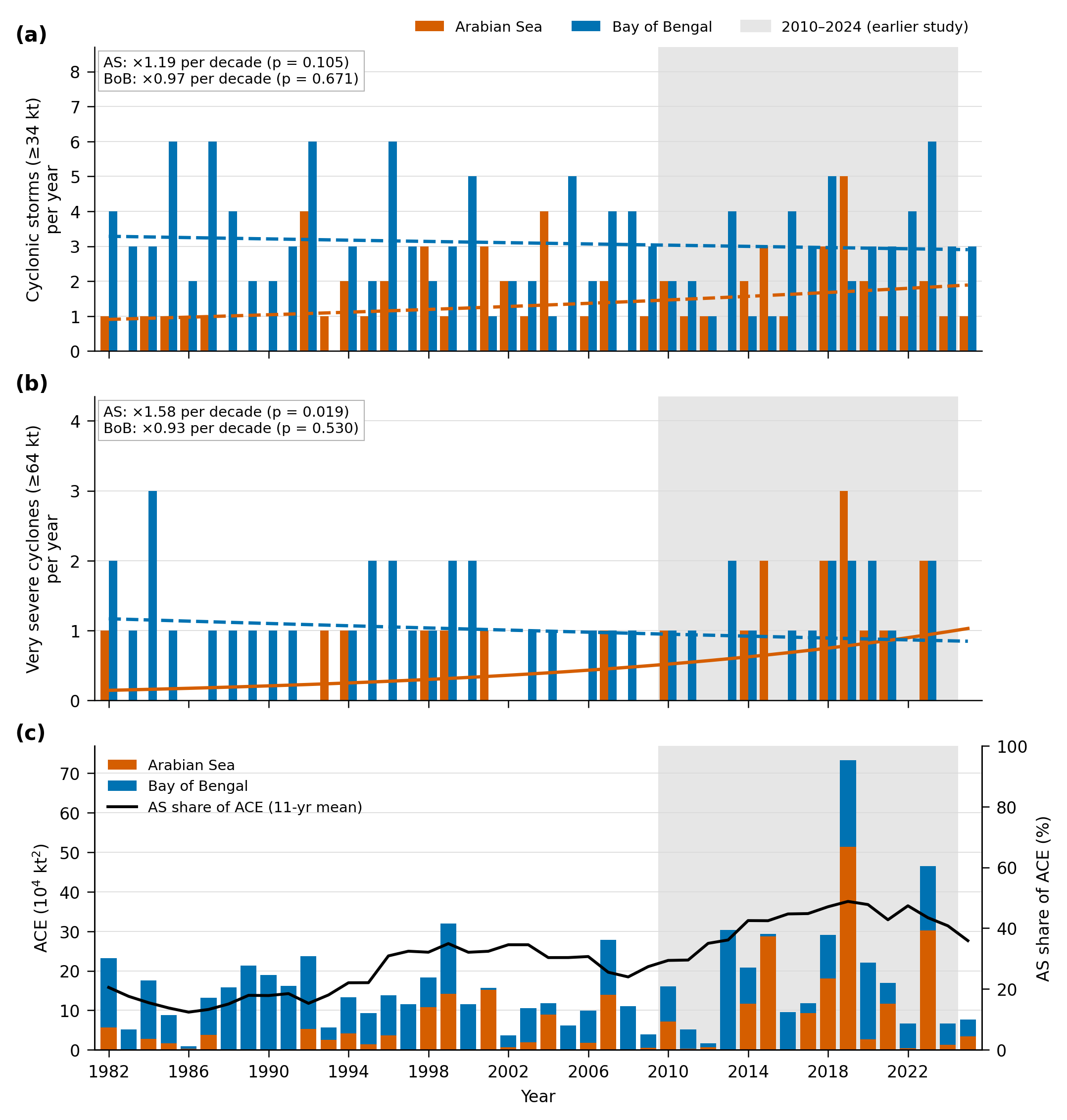}
\caption{Annual NIO TC activity, 1982--2025. (a) Cyclonic storms ($\geq$34\,kt) and (b) very severe cyclonic storms ($\geq$64\,kt) by sub-basin of genesis. Lines are negative-binomial GLM fits, solid where the trend is significant ($p<0.05$) and dashed otherwise; the rate ratio per decade and its $p$-value are given in each panel. (c) ACE by sub-basin (stacked bars) and the 11-year running AS share of NIO ACE (black line, right axis). Grey shading marks the 15-year window 2010--2024.}
\label{fig:timeseries}
\end{figure}

\begin{table}[t]
\centering
\caption{Trends in annual NIO TC metrics. Counts: negative-binomial GLM
rate ratio per decade [95\% CI]. ACE: Sen slope ($10^4$\,kt$^2$\,decade$^{-1}$)
with modified Mann--Kendall $p$. Bold: $p<0.05$. The last column gives the
count trend for 2010--2024 for comparison.}
\label{tab:trends}
\small
\begin{tabular}{@{}l l c c c@{}}
\toprule
\textbf{Metric} & \textbf{Basin} & \textbf{1982--2025} & $p$ & \textbf{2010--2024} ($p$) \\
\midrule
CS ($\geq$34\,kt) & AS  & 1.19 [0.97, 1.46] & 0.105 & 1.21 (0.68) \\
                        & BoB & 0.97 [0.85, 1.11] & 0.671 & 1.88 (0.08) \\
                        & NIO & 1.03 [0.92, 1.15] & 0.583 & 1.60 (0.10) \\
SCS ($\geq$48\,kt) & AS & \textbf{1.40 [1.04, 1.89]} & \textbf{0.027} & -- \\
                         & BoB & 0.98 [0.82, 1.18] & 0.862 & -- \\
VSCS ($\geq$64\,kt) & AS & \textbf{1.58 [1.08, 2.31]} & \textbf{0.019} & 1.73 (0.41) \\
                          & BoB & 0.93 [0.74, 1.17] & 0.530 & 1.11 (0.86) \\
RI events & AS & \textbf{1.98 [1.26, 3.12]} & \textbf{0.003} & -- \\
                & BoB & 0.86 [0.67, 1.10] & 0.234 & -- \\
ACE & AS & \textbf{+0.55} & \textbf{0.018} & -- \\
          & BoB & $-$1.00 & 0.307 & -- \\
          & NIO & +0.43 & 0.649 & -- \\
\midrule
\multicolumn{5}{@{}l}{Sensitivity, JTWC-only winds: AS VSCS 1.61 [1.15, 2.26], $p=0.005$;}\\
\multicolumn{5}{@{}l}{\phantom{Sensitivity, JTWC-only winds: }AS SCS 1.32 [1.03, 1.69], $p=0.031$.}\\
\bottomrule
\end{tabular}
\end{table}

Between 1982 and 2025, 416 systems that formed in the NIO were analysed, 102 in the AS and 314 in the BoB. Of these, 195 reached cyclonic-storm (CS) strength and 64 became very severe cyclonic storms (VSCS) or stronger. The number of cyclonic storms in the basin as a whole has not changed: the rate ratio is
1.03\,decade$^{-1}$ ($p=0.58$; Table~\ref{tab:trends};
Fig.~\ref{fig:timeseries}a). What has changed is the strength of the storms that form in the AS. The number of AS systems of all strengths also rose (1.19 per decade, $p=0.033$), but the trends grow with intensity: severe and very severe cyclonic storms there increased by factors of 1.40 and 1.58 per decade, RI events by 1.98 per decade, and AS ACE by $0.55\times10^4$\,kt$^2$ per decade ($p=0.018$). No BoB metric shows a significant trend in either direction. The AS increase is not an artefact of merging the two agencies' winds. With JTWC 1-min winds alone, the AS VSCS trend is 1.61\,decade$^{-1}$ ($p=0.005$) and the SCS trend 1.32\,decade$^{-1}$ ($p=0.031$).

Splitting the record into two halves illustrates the size of the change, which is tested by the trends above. The AS share of NIO ACE rose from 24\% in 1982--2003 to 50\% in 2004--2025, and its share of post-monsoon (OND) ACE from 16\% to 51\% (Fig.~\ref{fig:timeseries}c). The number of AS storms that reached VSCS more than doubled, from 6 to 14, while the BoB number hardly changed (24 and 20).
The basin as a whole has therefore not become more active, but the AS now produces about half of its cyclone energy. Neither the CS nor the VSCS trend is significant over 2010–2024 alone (Table~\ref{tab:trends}, last column), which is the subject of the next section.

\subsection{How long a record is needed to detect a trend}
\label{subsec:window}

\begin{figure}[t]
\centering
\includegraphics[width=\textwidth]{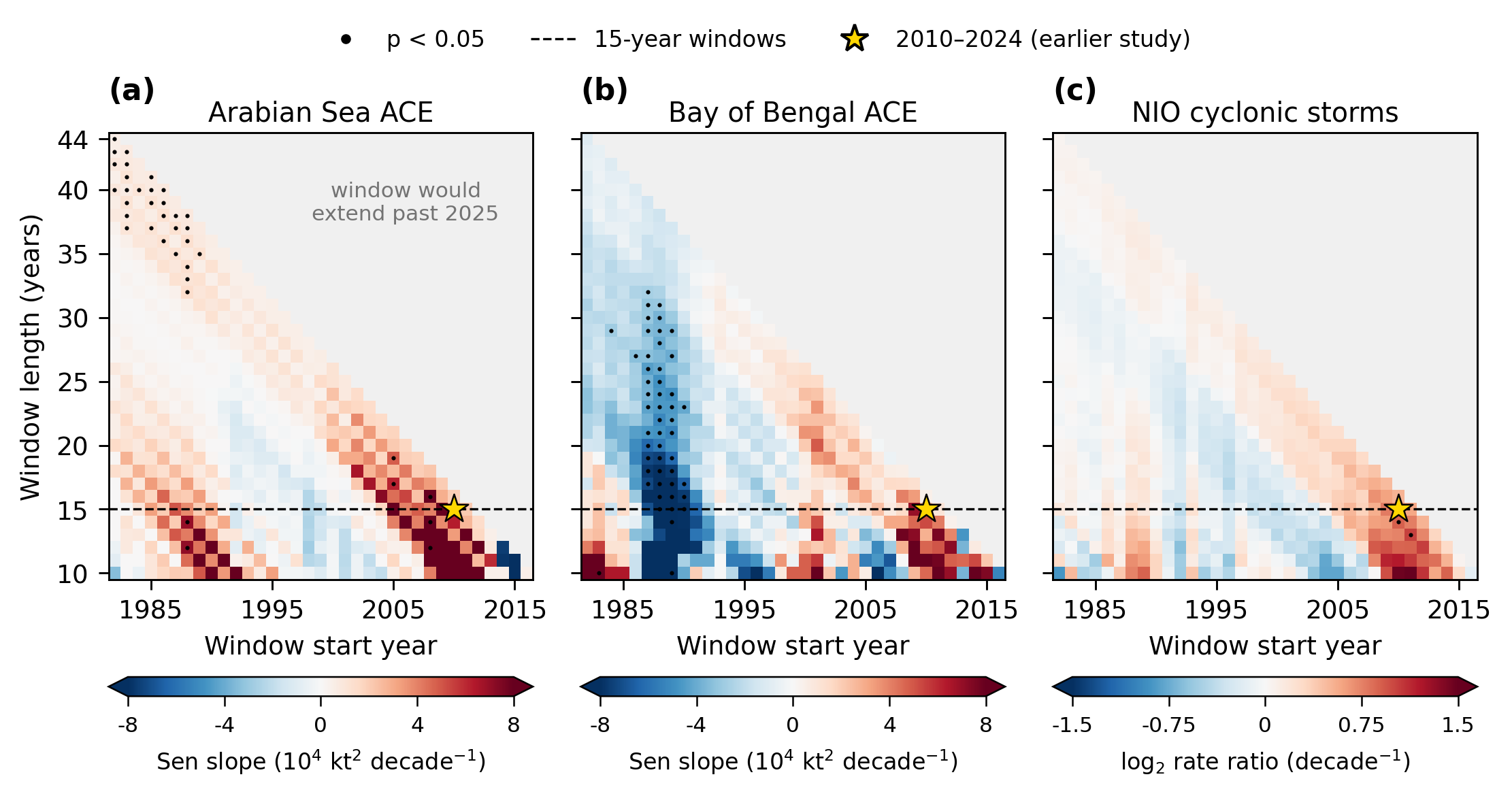}
\caption{Dependence of the trend on record length and start year within 1982--2025. Each cell is the trend over a window of the given length (y-axis) starting in the given year (x-axis) for (a) AS ACE and (b) BoB ACE (Sen slope, $10^4$\,kt$^2$\,decade$^{-1}$) and (c) NIO cyclonic-storm counts (log$_2$ of the Poisson GLM rate ratio per decade). Black dots mark windows with $p<0.05$. The grey area contains windows that would extend beyond 2025. The dashed line marks 15-year windows and the star the 2010--2024 window.}
\label{fig:window}
\end{figure}

To see how much a trend depends on the period chosen, we recomputed it for every window that fits inside 1982--2025 (Fig.~\ref{fig:window}). For AS ACE, 15-year windows give trends of both signs (60\% of them positive), and none of the 30 possible 15-year windows has a significant trend, even though the 44-year trend is significant ($p=0.018$). The few short windows that are significant (12--19 years long, starting in 1988 or in 2005--2008) give slopes of 4.9--18.7$\times10^4$\,kt$^2$ per decade, 9 to 34 times the
long-term value, because they contain the extreme seasons of 2015 and 2019. A significant AS increase appears consistently only in windows of 32 years or more that start in the 1980s (Fig.~\ref{fig:window}a).

The BoB shows the opposite problem. Three of the 30 fifteen-year windows (those starting in 1988--1990) have significant \emph{negative} trends, and negative trends of $-3$ to $-14\times10^4$\,kt$^2$ per decade remain significant in windows of up to 32 years that start in 1984--1990. Over the
full record there is no BoB trend at all ($p=0.31$). For NIO cyclonic-storm counts, windows starting after about 2005 give large positive rate ratios that disappear in longer records. The 2010--2024 window is one of them (star in Fig.~\ref{fig:window}c; 1.60\,decade$^{-1}$, $p=0.10$). A 15-year record therefore tells us more about the decadal phase it happens to sample than about long-term change. Trends in NIO TC activity, or claims that one sub-basin has become dominant, need at least about three decades of data.

\subsection{ENSO redistributes post-monsoon cyclone energy between the two basins}
\label{subsec:redistribution}

\begin{figure}[t]
\centering
\includegraphics[width=\textwidth]{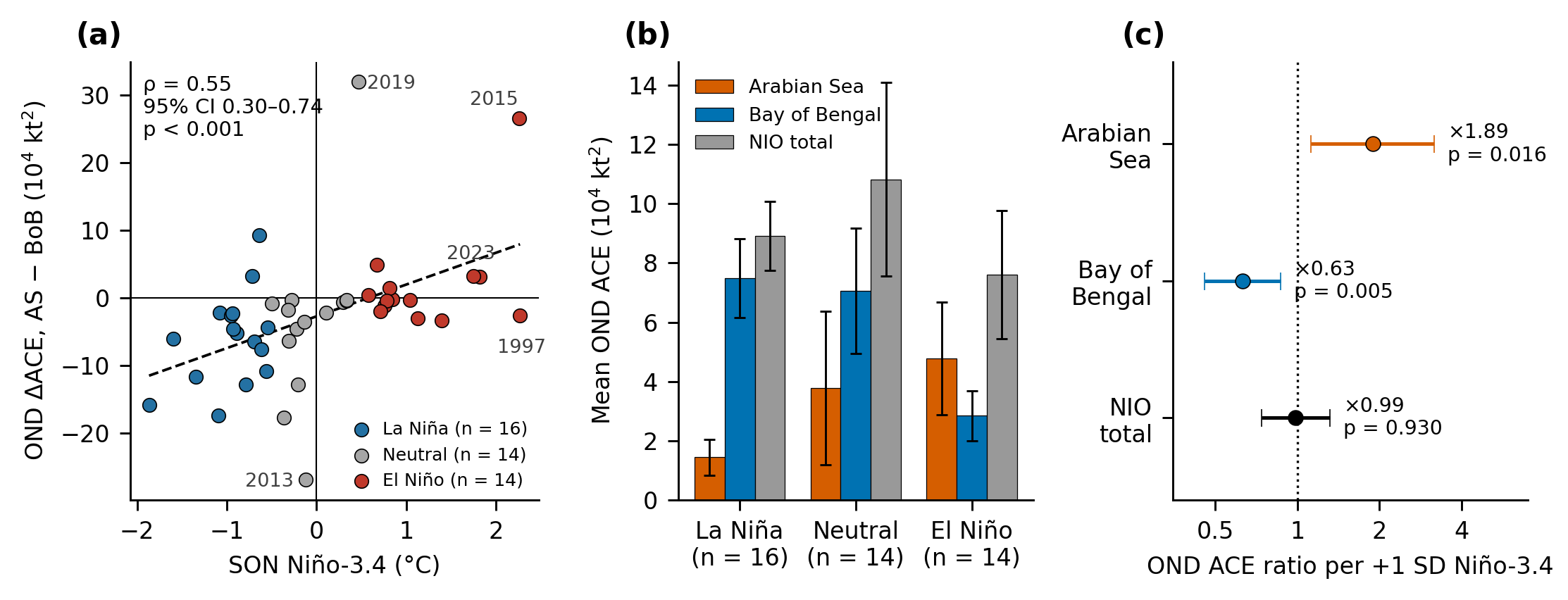}
\caption{ENSO and OND TC energy, 1982--2025. (a) $\Delta\ACE$
(AS minus BoB) against SON Ni\~no-3.4 (red: El Ni\~no, blue: La Ni\~na, grey: neutral). (b) Mean OND ACE ($\pm$1 s.e.) by ENSO phase. (c) Quasi-Poisson GLM ACE ratio per +1 SD of Ni\~no-3.4, with 95\% CIs.}
\label{fig:redistribution}
\end{figure}

\begin{table}[htbp]
\centering
\caption{Robustness of the Spearman correlation between OND $\Delta \ACE$
and SON Ni\~no-3.4 (baseline $\rho=0.55$, permutation $p<0.001$). All variants have $p\leq0.01$.}
\label{tab:robust}
\small
\begin{tabular}{@{}l c@{}}
\toprule
\textbf{Variant} & $\rho$ \\
\midrule
Baseline (merged winds, fix-based allocation) & 0.55 \\
IMD winds only & 0.53 \\
JTWC 1-min winds only & 0.54 \\
Storm-based (genesis) basin allocation & 0.65 \\
Both series linearly detrended & 0.55 \\
Leave-one-year-out (range) & 0.52--0.60 \\
Two largest AS seasons removed (2015, 2019) & 0.51 \\
Provisional 2025 tracks excluded & 0.53 \\
Sub-period 1982--2003 / 2004--2025 & 0.65 / 0.54 \\
Pearson instead of Spearman & 0.48 \\
Moving-block bootstrap 95\% CI (3-yr blocks) & [0.33, 0.72] \\
\bottomrule
\end{tabular}
\end{table}

If ENSO simply raised or lowered NIO activity, the AS and the BoB would move together. They do not. In OND, the energy difference between the two basins rises with SON Ni\~no-3.4 ($\rho=0.55$ [0.30, 0.74], $p<0.001$; Fig.~\ref{fig:redistribution}a), while total NIO OND ACE has only a weak and non-significant negative relationship with it ($\rho=-0.24$, $p=0.12$). The GLM, which works with magnitudes, gives the same picture (Fig.~\ref{fig:redistribution}c). A one-standard-deviation warming of Ni\~no-3.4 multiplies AS OND ACE by 1.89 [1.13, 3.18] ($p=0.016$) and BoB OND ACE by 0.63 [0.46, 0.87] ($p=0.005$). The two effects almost cancel, and the ratio for the NIO total is 0.99 [0.74, 1.32] ($p=0.93$). Over 1982--2025, El Ni~no has moved post-monsoon cyclone energy from the BoB to the AS without a detectable change in the NIO total; the interval for the total rules out changes larger than about 25--30\% per standard deviation.

This result does not depend on a particular data choice
(Table~\ref{tab:robust}). In all eleven tests the correlation stays positive and significant. The Spearman variants lie between 0.51 and 0.65; they include single-agency winds, genesis-based allocation of storms to basins, detrending, leave-one-year-out resampling, removal of 2015 (Chapala and Megh) and 2019 (Kyarr and Maha), exclusion of the provisional 2025 tracks, and each half of the record taken separately. The Pearson correlation, which gives more weight to the few very large AS seasons, is 0.48, and the moving-block
bootstrap interval [0.33, 0.72] does not include zero. In rank terms the BoB side of the dipole is the stronger one: BoB OND ACE correlates with Ni\~no-3.4 at $\rho=-0.47$ ($p=0.001$), against $+0.23$ ($p=0.14$) for the
AS. This is partly because 15 of the 44 OND seasons had no AS ACE at all, and these tied zeros weaken a rank correlation. The GLM, which uses the size of ACE rather than its rank, does detect the AS increase.

\subsection{ENSO acts on intensification rather than formation}
\label{subsec:intensification}

\begin{figure}[t]
\centering
\includegraphics[width=\textwidth]{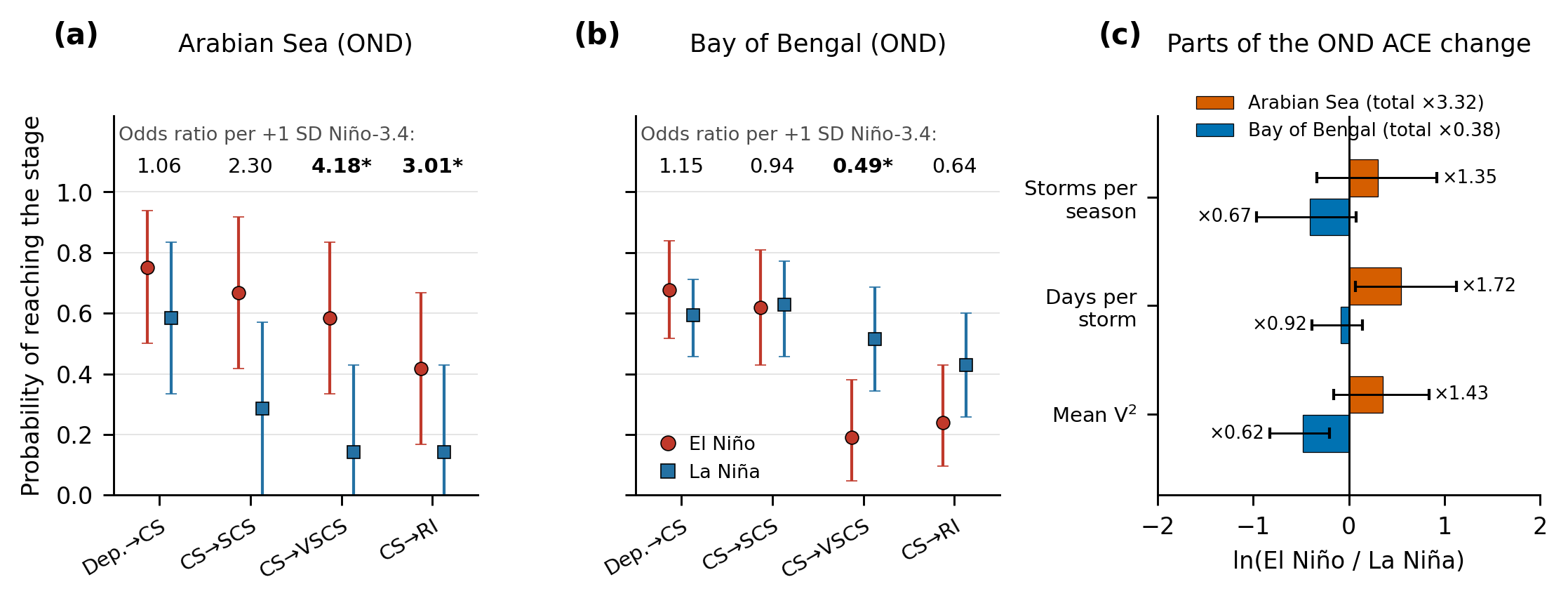}
\caption{Stage of the TC life cycle affected by ENSO in OND. (a, b) Conditional probabilities in El Ni\~no (red) and La Ni\~na (blue) seasons (bootstrap 95\% CIs) for the AS and BoB. Odds ratios (OR) are per +1 SD of SON Ni\~no-3.4 from logistic regression on all seasons; $^*p<0.05$. (c)
Decomposition of the El Ni\~no--La Ni\~na difference in $\ln \ACE$ into storm frequency, TC-days per storm and mean squared wind (bootstrap 95\% CIs).}
\label{fig:intensification}
\end{figure}

To find out which stage of a storm's life ENSO acts on, we separated two steps: a D becoming a CS, and a CS becoming
a VSCS (Fig.~\ref{fig:intensification}a,b). ENSO has no detectable effect on the first step in either basin: the odds ratio per standard deviation of Ni~no-3.4 is 1.06 [0.59, 1.91] ($p=0.84$) in the AS and 1.15 [0.79, 1.67] ($p=0.47$) in the BoB. The second step responds strongly, and in opposite directions: the odds ratio is 4.18 [1.27, 13.7] ($p=0.018$) in the AS and 0.49 [0.27, 0.89] ($p=0.018$) in the BoB. In plain numbers, 58\% of AS cyclonic storms (7 of 12) reached VSCS in El Ni~no seasons compared with 14\% (1 of 7) in La Ni~na seasons; in the BoB the figures were 19\% (4 of 21) and 51\% (18 of 35). RI shows the same pattern, significantly in the AS (OR 3.01, $p=0.047$) and more weakly in the BoB (OR 0.64, $p=0.11$). El Ni\~no does reduce the number of depressions that form in the BoB (rate ratio 0.81\,SD$^{-1}$, $p=0.020$). Because the chance of a D becoming a storm is unchanged, however, the effect on the number of BoB cyclonic storms is small and not significant (0.85, $p=0.17$).

Splitting OND ACE into storm number, storm duration and mean squared wind (Section~\ref{subsec:statistical_analysis}) shows that the two basins reach their response by different routes (Fig.~\ref{fig:intensification}c). In the AS, most of the El Ni\~no--La Ni\~na difference in $\ln \ACE$ comes from
duration ($\Delta\ln D=+0.54$ [0.08, 1.08], i.e. storms lasting about 1.7 times as long), with smaller and uncertain contributions from frequency ($+0.30$) and intensity ($+0.36$). In the BoB, ACE falls mainly because storms are weaker ($\Delta\ln\overline{v^2}=-0.48$ [$-0.81$, $-0.19$], a
factor of 0.62) and somewhat fewer ($-0.41$), while their duration hardly changes. El Ni\~no therefore lets AS storms live longer and keeps BoB storms weaker. In neither basin does it significantly change the number of cyclonic storms.

\subsection{Spatial patterns}
\label{subsec:spatial}

\begin{figure}[t]
\centering
\includegraphics[width=\textwidth]{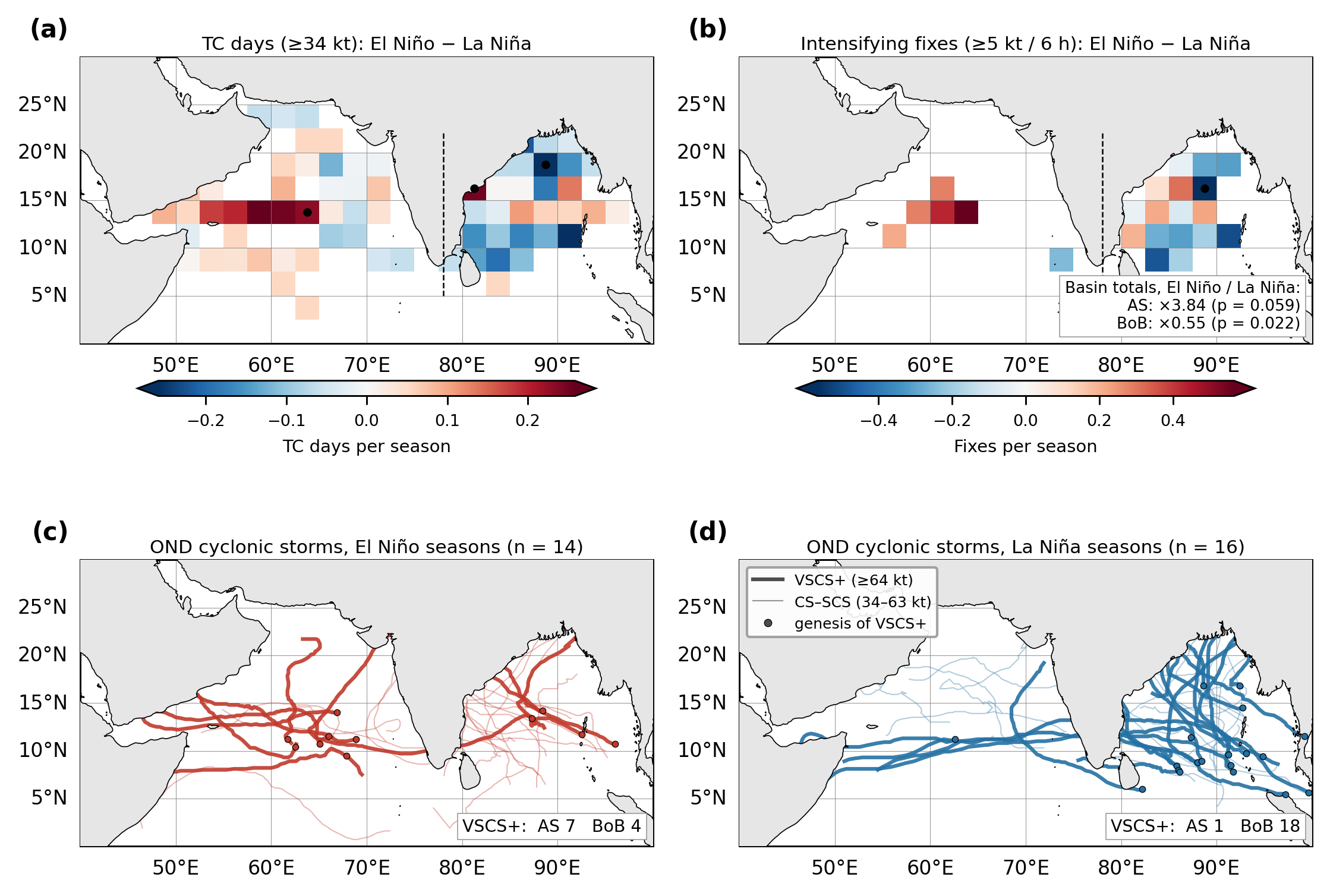}
\caption{(a) El Ni\~no minus La Ni\~na difference in OND TC-days ($\geq$34\,kt) per season and (b) intensifying fixes ($\geq$5\,kt per 6\,h) per season, on a 2.5$^\circ$ grid. Black dots mark cells with pointwise permutation $p<0.05$; no cell passes FDR control at $q=0.10$. The dashed line near 78$^\circ$E separates the two sub-basins, and the box in (b) gives the El Ni\~no/La Ni\~na ratio of basin-total intensifying fixes with its Mann--Whitney $p$. (c, d) Tracks of OND cyclonic storms in El Ni\~no (14 seasons) and La Ni\~na (16 seasons) seasons; thick lines and dots (genesis
positions) mark storms that reached VSCS.}
\label{fig:enso_maps}
\end{figure}

The storm tracks make the redistribution easy to see
(Fig.~\ref{fig:enso_maps}c,d). In the 14 El Ni\~no seasons, 7 AS and 4 BoB storms reached VSCS. In the 16 La Ni\~na seasons only one AS storm did, against 18 in the BoB. Summed over each basin, intensifying fixes were 3.8 times as frequent in the AS in El Ni\~no seasons ($p=0.059$) and about half
as frequent in the BoB ($\times0.55$, $p=0.022$). On the maps, the El Ni\~no increase lies mainly in the central AS (10--16$^\circ$N, 55--65$^\circ$E) and the decrease in the southern and central Bay (Fig.~\ref{fig:enso_maps}a,b). Individual grid cells contain few storms, though: only three cells for TC-days and one for intensifying fixes pass a
pointwise test, and none survives FDR control. Our conclusions therefore rest on the basin-integrated and storm-level tests, and the maps are used only to show where the changes occur.

\begin{figure}[t]
\centering
\includegraphics[width=\textwidth]{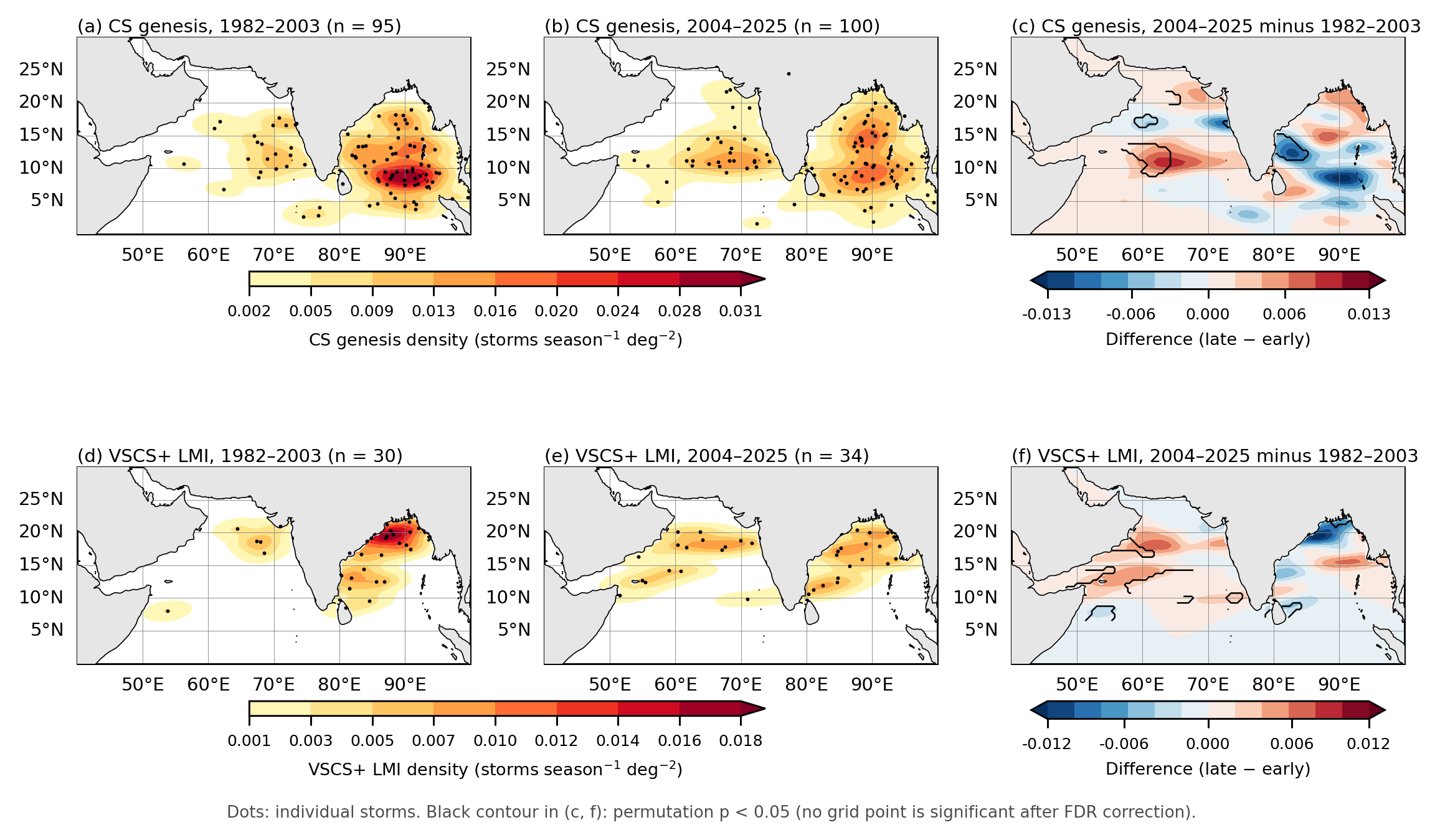}
\caption{Kernel density of (a--c) cyclonic-storm genesis following \cite{wahiduzzaman2017climatological} and (d--f) VSCS+ LMI locations for 1982--2003, 2004--2025 and their difference (per season). Black contours in (c) and (f) enclose sea points with storm activity where the pointwise $p<0.05$, from 10,000 permutations of whole years between the two periods. No point remains significant after FDR control at $q=0.10$.}
\label{fig:kde}
\end{figure}

The same care is needed for changes in where storms form. Fig.~\ref{fig:kde} compares the density of cyclonic-storm genesis and of VSCS+ LMI positions between 1982--2003 and 2004--2025. In the later period there is more genesis in the central AS and less in the southern BoB, and the number of AS
cyclonic storms forming west of 65$^\circ$E rose from 4 to 9. The LMI density of the strongest storms moved from the northern BoB towards the central AS (Fig.~\ref{fig:kde}f), in line with the AS intensity trend. None of these differences is field-significant. Although some sea points with storm activity pass the pointwise test, none remains significant after FDR control. The storm positions tell the same story. The mean genesis longitude of AS cyclonic storms ($-0.62^\circ$\,decade$^{-1}$, $p=0.32$), the mean LMI
longitude of AS VSCS ($-1.60^\circ$\,decade$^{-1}$, $p=0.53$) and the mean genesis latitude of BoB cyclonic storms ($p=0.69$) have no significant trend. Over 1982--2025, the robust change is in how strong AS storms become and how much energy they carry (Section~\ref{subsec:trends}), not in where
storms form.

\subsection{Separate roles of ENSO and the IOD by season}
\label{subsec:iod}

\begin{figure}[t]
\centering
\includegraphics[width=\textwidth]{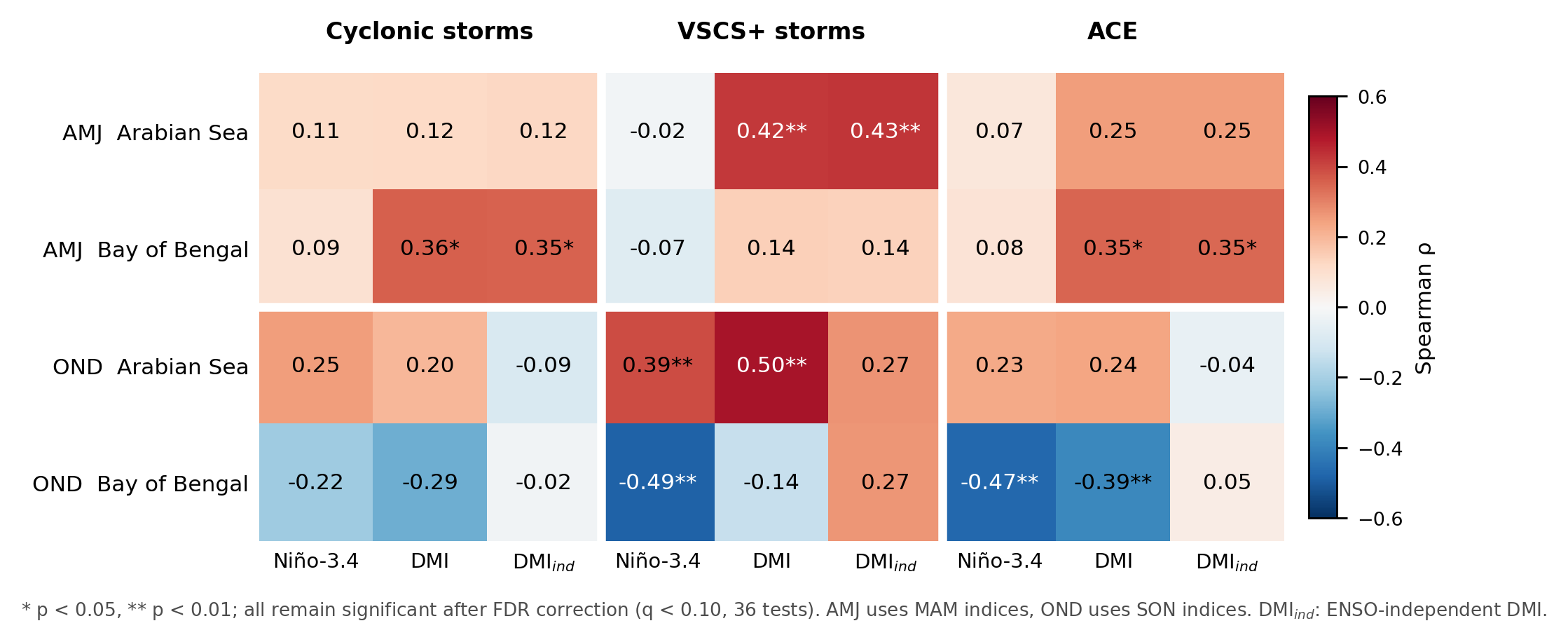}
\caption{Spearman correlation of seasonal TC metrics with Ni\~no-3.4, DMI and ENSO-independent DMI (DMI$_{ind}$) for AMJ (MAM indices) and OND (SON indices), 1982--2025. $^*p<0.05$, $^{**}p<0.01$. All significant correlations remain significant after Benjamini--Hochberg FDR control at
$q=0.10$.}
\label{fig:matrix}
\end{figure}

\begin{table}[htbp]
\centering
\caption{Mean OND ACE ($10^4$\,kt$^2$) by combined ENSO--IOD phase,
1982--2025 (SON indices; ENSO $\pm$0.5\,\degC, IOD $\pm$0.4\,\degC). The
ENSO$\times$IOD interaction is not significant (GLM $p=0.12$ for the AS,
$p=0.59$ for the BoB). The single neutral-ENSO, positive-IOD season is
2019.}
\label{tab:enso_iod}
\small
\begin{tabular}{@{}l l c c c c@{}}
\toprule
\textbf{ENSO} & \textbf{IOD} & $n$ & \textbf{AS} & \textbf{BoB} & $\Delta ACE$ \\
\midrule
El Ni\~no & positive & 6 & 4.02 & 3.51 & +0.51 \\
El Ni\~no & neutral  & 8 & 5.33 & 2.34 & +2.99 \\
Neutral   & positive & 1 & 37.15 & 5.21 & +31.94 \\
Neutral   & neutral  & 10 & 0.89 & 6.33 & $-$5.44 \\
Neutral   & negative & 3 & 2.23 & 10.10 & $-$7.87 \\
La Ni\~na & neutral  & 13 & .33 & 7.45 & $-$6.13 \\
La Ni\~na & negative & 3 & 1.90 & 7.63 & $-$5.73 \\
\bottomrule
\end{tabular}
\end{table}

In the post-monsoon season the IOD is hard to separate from ENSO. SON Ni\~no-3.4 and the DMI are correlated at $r=0.66$, so much of the apparent IOD effect repeats the ENSO effect. OND $\Delta \ACE$ correlates with the DMI at $\rho=0.54$, but with the ENSO-independent DMI at only $\rho=0.09$ ($p=0.55$; Fig.~\ref{fig:matrix}). Each index remains marginally significant when the other is held fixed (partial $\rho=0.33$ and 0.31, $p\approx0.03$ and 0.04), so 44 seasons are not enough to separate the two cleanly. The ENSO$\times$IOD interaction is not significant either (Table~\ref{tab:enso_iod}). The one clear case of an IOD-driven season is 2019. It was one of the three strongest positive IOD events in the record (SON DMI $=0.90$\,\degC) and the only one of them that occurred with
near-neutral ENSO (SON Ni\~no-3.4 $=0.47$\,\degC), and it produced the largest AS OND ACE of the 44 years ($37.2\times10^4$\,kt$^2$). A quasi-Poisson GLM picks up an AS response to DMI$_{\mathrm{ind}}$ (ratio 1.80, $p=0.03$), but the rank correlation does not ($\rho=-0.04$), which
suggests that the GLM result rests largely on this one season. In OND, then, the IOD signal is either carried by ENSO or limited to rare seasons in which the two modes are decoupled. Table~\ref{tab:enso_iod} gives the phase composites for completeness; with one to thirteen seasons per class, they should not be read as a scheme for prediction.

The pre-monsoon season behaves differently. In AMJ, ENSO has no detectable relationship with any TC metric (Fig.~\ref{fig:matrix}). The IOD, which in MAM is almost independent of ENSO, is associated with more AS VSCS
($\rho=0.42$; DMI$_{\mathrm{ind}}$: 0.43) and with higher BoB cyclonic-storm counts and ACE ($\rho=0.35$--0.36). At the storm level, a one-standard-deviation increase in DMI$_{\mathrm{ind}}$ raises the odds that an AMJ AS cyclonic
storm becomes a VSCS by a factor of 4.9 (95\% CI 1.3--19.0, $p=0.022$); the fitted probability rises from about 7\% at $-1$\,SD to 65\% at $+1$\,SD. Only 24 storms are available, so this estimate is uncertain. Of the 36 correlations in Fig.~\ref{fig:matrix}, 11 are significant at $p<0.05$, and
all of them survive FDR control ($q\leq0.07$). ENSO thus controls the post-monsoon season, whereas an influence of the IOD that is independent of ENSO appears in the pre-monsoon season.

\subsection{A change in the ENSO--TC relationship over time}
\label{subsec:nonstationary}

\begin{figure}[t]
\centering
\includegraphics[width=\textwidth]{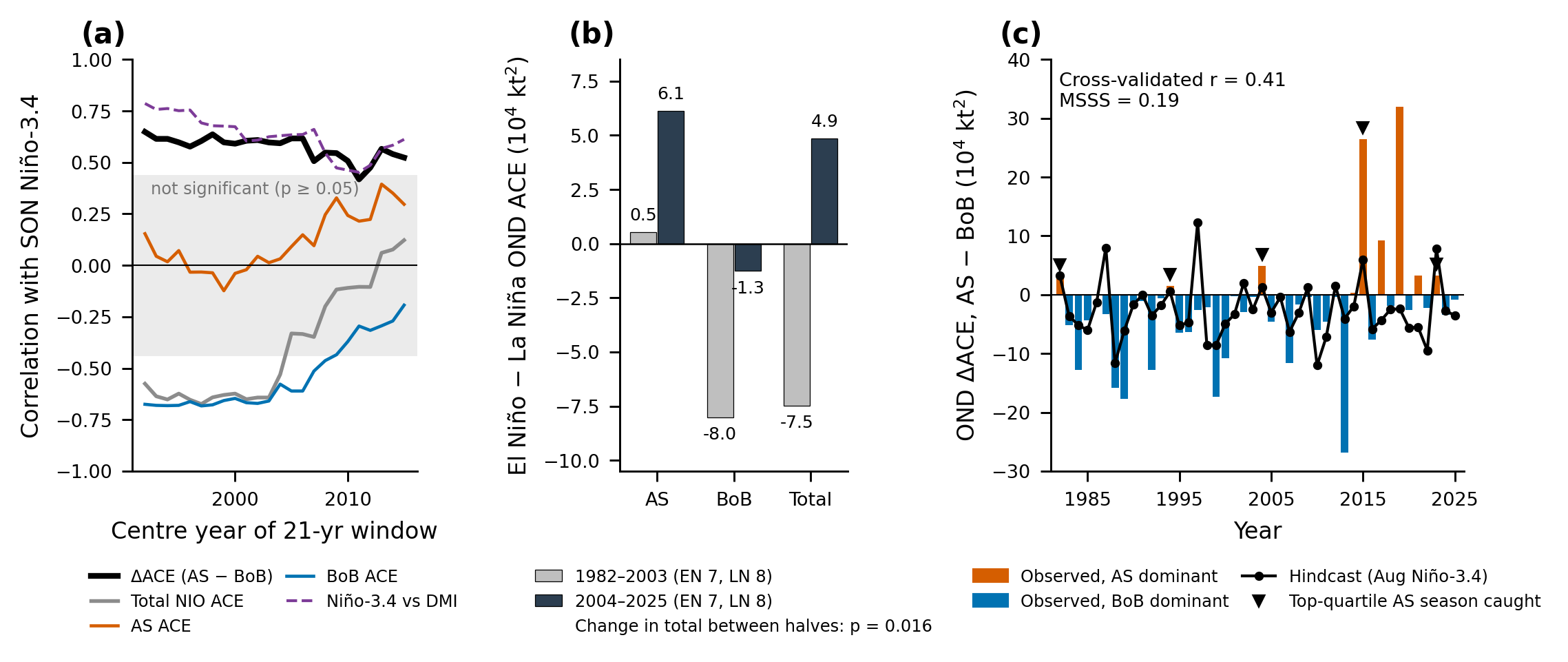}
\caption{(a) 21-year running Spearman correlations of SON Ni\~no-3.4 with OND $\Delta ACE$, total NIO ACE, BoB ACE and AS ACE, and the running Ni\~no-3.4--DMI correlation (dashed), plotted at the centre year of each window. Correlations inside the grey band ($|\rho|<0.44$) are not significant at $p<0.05$ for $n=21$. (b) El Ni\~no minus La Ni\~na OND ACE in
the AS, the BoB and the NIO total for 1982--2003 and 2004--2025; the legend gives the number of El Ni\~no and La Ni\~na seasons in each half and the permutation $p$ for the change in the total. (c) Observed OND $\Delta ACE$ (bars, coloured by the dominant basin) and its leave-one-out hindcast from
August Ni\~no-3.4 (black line); triangles mark top-quartile AS-dominated seasons that the hindcast also placed in its top quartile.}
\label{fig:stationarity}
\end{figure}

The link between ENSO and $\Delta \ACE$ is stable through time: its 21-year running correlation stays positive throughout (0.42--0.65) and is significant in all but one of the 24 windows (Fig.~\ref{fig:stationarity}a). What has changed is how the two basins contribute to it. In the first half of the record (1982--2003), the BoB response dominated ($\rho=-0.67$, $p=0.001$) and the AS response was weak
($\rho=0.15$). El Ni\~no therefore lowered the \emph{total} post-monsoon energy of the NIO ($\rho=-0.59$, $p=0.004$); in that period ENSO modulated the basin. In 2004--2025, the BoB response weakened ($\rho=-0.25$), the AS response strengthened ($\rho=0.34$), and the relationship with total energy disappeared ($\rho=+0.08$, $p=0.72$). The change in the total-energy correlation is significant (permutation $p=0.016$; Fisher $z$ test $p=0.022$), and it does not depend on where the record is split. For any split year between 1996 and 2012, the early-period correlation lies between $-0.48$ and $-0.63$ and the late-period correlation between $-0.12$ and $+0.60$. Leaving out one year at a time moves the late-period value only between $-0.04$ and $+0.20$.

The phase composites show the same shift (Fig.~\ref{fig:stationarity}b). In 1982--2003, El Ni\~no seasons had $7.5\times10^4$\,kt$^2$ less total OND ACE
than La Ni\~na seasons, and almost all of the difference came from the BoB ($-8.0$, against $+0.5$ in the AS). In 2004--2025 they had $4.9\times10^4$\,kt$^2$ \emph{more}, because the AS gain ($+6.1$) outweighed a much smaller BoB loss ($-1.3$). The relationship has thus moved from modulation of the basin total towards redistribution between the two sub-basins. A study covering only the earlier decades would find that El Ni\~no suppresses NIO activity, and one covering only the recent decades would find little effect on the total; this change in time may account for part of the disagreement among published estimates (Section~\ref{sec:discussion}). Over the same period the SON
Ni\~no-3.4--DMI correlation fell from 0.79 in the first window to a minimum of 0.45 in the window centred on 2011, before partly recovering (0.61 for 2005--2025; Fig.~\ref{fig:stationarity}a). A weakening of the ENSO influence on the Indian Ocean since the early 2000s has been reported \citep{yu2025weakened}, but our data cannot show whether the two changes are connected.

\subsection{Environmental pathways}
\label{subsec:env}

\begin{figure}[t]
\centering
\includegraphics[width=\textwidth]{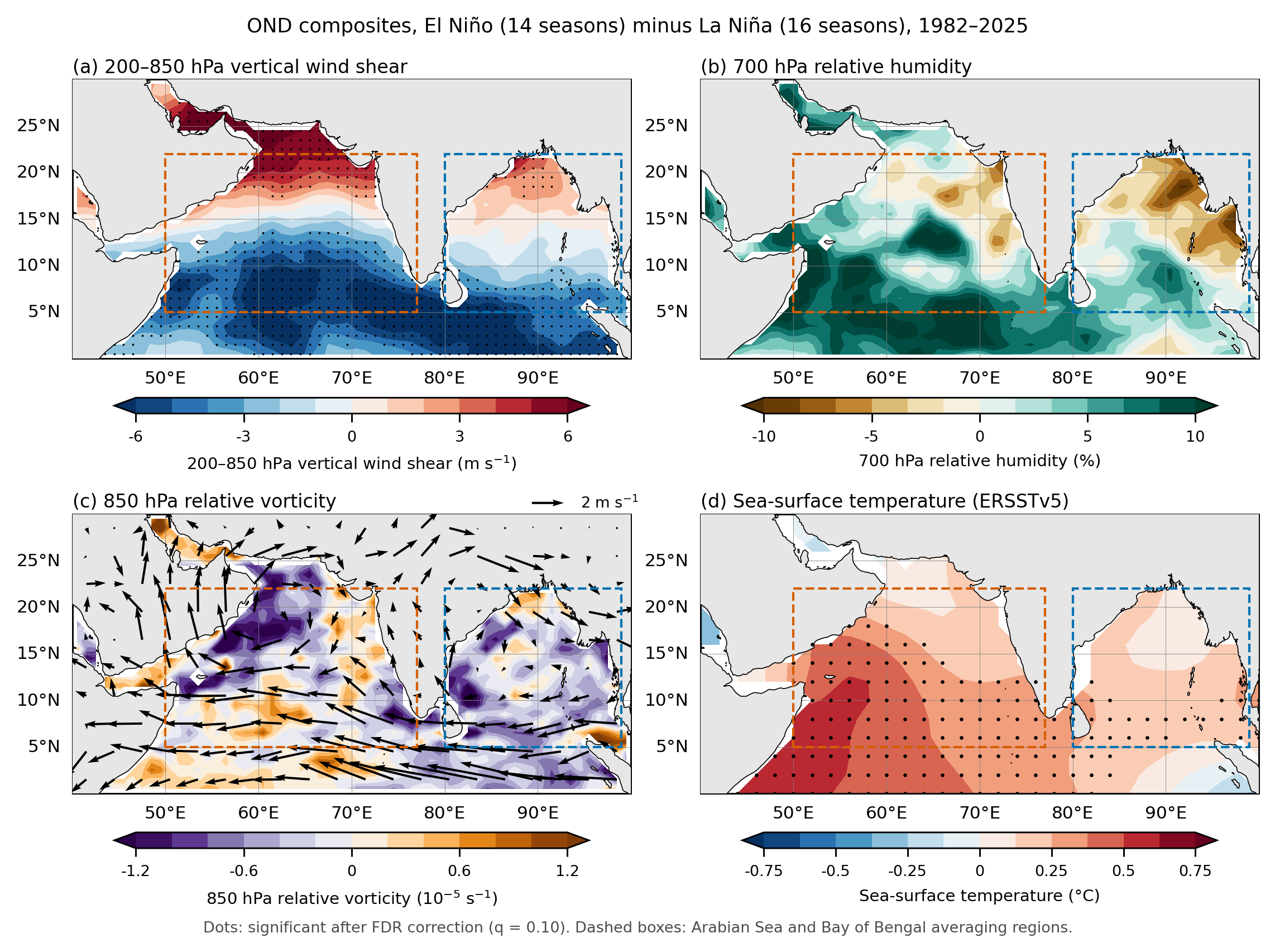}
\caption{OND El Ni\~no minus La Ni\~na composites: (a) 200--850\,hPa VWS, (b) 700\,hPa RH, (c) 850\,hPa relative vorticity with 850\,hPa wind anomalies (vectors), from ERA5, and (d) SST from ERSSTv5, all for 1982--2025. Dots: FDR-significant at $q=0.10$. Dashed boxes are the AS and BoB averaging regions.}
\label{fig:env}
\end{figure}

\begin{figure}[t]
\centering
\includegraphics[width=\textwidth]{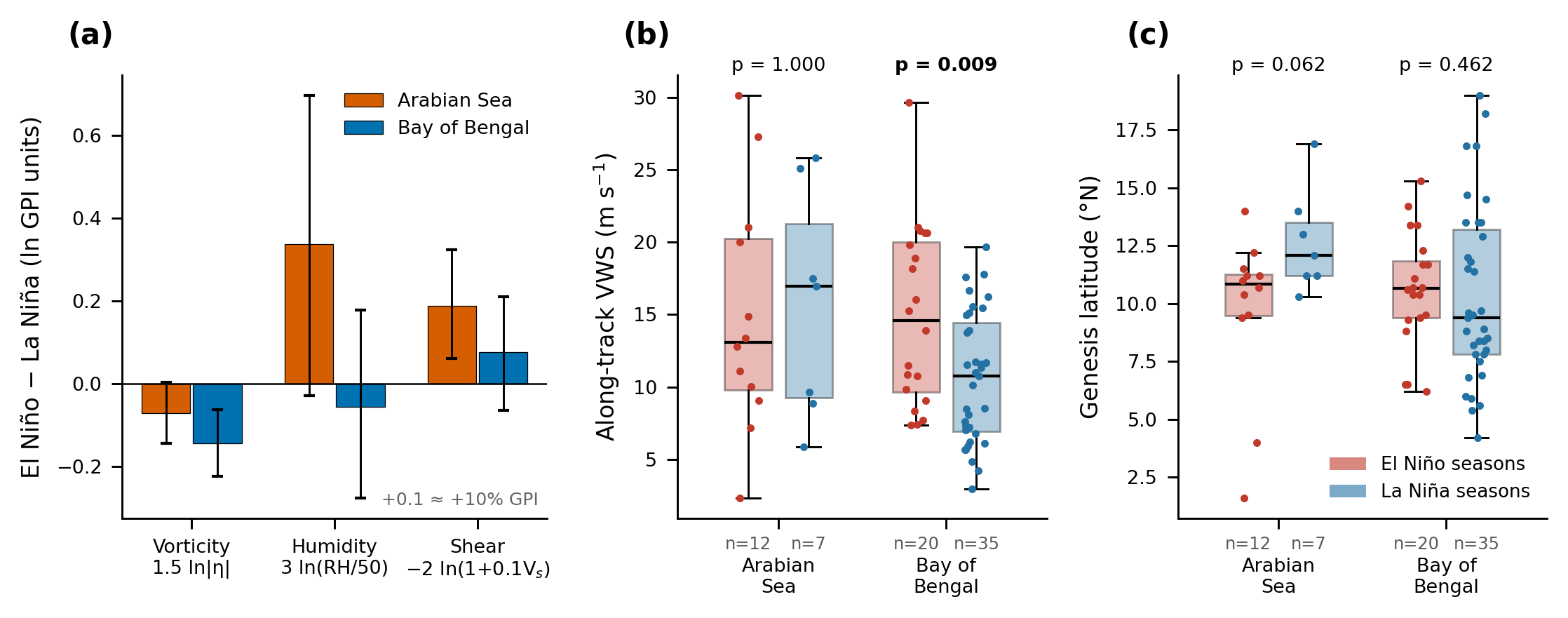}
\caption{(a) El Ni\~no minus La Ni\~na difference in the atmospheric terms of the genesis potential index \citep{camargo2007use}, averaged over the AS and BoB boxes (bootstrap 95\% CIs). (b) VWS averaged along the track of each OND cyclonic storm, from the time it first reached 34\,kt to its LMI, in El Ni\~no and La Ni\~na seasons. (c) Genesis latitude of the same storms. $p$-values from Mann--Whitney tests.}
\label{fig:mech}
\end{figure}

\begin{table}[htbp]
\centering
\caption{OND box-mean environmental response to ENSO. EN$-$LN: El Ni\~no minus La Ni\~na mean (Mann--Whitney $p$); $\rho$: Spearman correlation with SON Ni\~no-3.4, for 1982--2003 and 2004--2025.}
\label{tab:env}
\small
\begin{tabular}{@{}l l c c c c@{}}
\toprule
\textbf{Variable} & \textbf{Box} & \textbf{EN$-$LN} & $p$ & $\rho$ (1982--2003) & $\rho$ (2004--2025) \\
\midrule
VWS (m\,s$^{-1}$) & AS & \textbf{$-$2.29} & \textbf{0.008} & $-$0.53 & $-$0.37 \\
 & BoB & $-$0.91 & 0.25 & $-$0.27 & $-$0.16 \\
RH$_{700}$ (\%) & AS & +3.65 & 0.11 & 0.36 & 0.26 \\
 & BoB & $-$0.66 & 0.47 & $-$0.01 & $-$0.08 \\
$\zeta_{850}$ ($10^{-5}$\,s$^{-1}$) & AS & \textbf{$-$0.21} & \textbf{0.006} & $-$0.58 & $-$0.36 \\
 & BoB & \textbf{$-$0.31} & \textbf{0.001} & $-$0.54 & $-$0.42 \\
SST (\degC) & AS & \textbf{+0.32} & \textbf{0.006} & -- & -- \\
 & BoB & +0.16 & 0.27 & -- & -- \\
\bottomrule
\end{tabular}
\end{table}

ENSO changes the environment of the two basins in different ways (Fig.~\ref{fig:env}; Table~\ref{tab:env}). In El Ni\~no seasons, easterly anomalies at 850\,hPa extend across the equatorial Indian Ocean (Fig.~\ref{fig:env}c). VWS forms a north--south dipole in both basins, lower south of about 15$^\circ$N and higher to the north, with FDR-significant points over large areas (Fig.~\ref{fig:env}a). Averaged over
the AS box, shear falls by 2.3\,m\,s$^{-1}$ ($p=0.008$). The AS box is also warmer (+0.32\,\degC, $p=0.006$) and, less clearly, moister (+3.6\% RH$_{700}$, $p=0.11$). Because the AS warms more than the BoB, the AS-minus-BoB OND SST difference also increases with Ni\~no-3.4 ($\rho=0.31$,
$p=0.039$). In the BoB box, mean shear and humidity change little. The clearest BoB signal is weaker cyclonic vorticity at 850\,hPa ($-0.31\times10^{-5}$\,s$^{-1}$, $p=0.001$), and this happens even though BoB SST is slightly higher. The BoB suppression is therefore dynamical rather than thermodynamic.

The genesis potential terms bring this contrast out more clearly (Fig.~\ref{fig:mech}a). In the AS, the shear term is the only significant positive contribution (+0.19 [0.06, 0.32] in ln units), helped by a positive but uncertain humidity term (+0.34). In the BoB, the only significant term
is absolute vorticity, and it is negative ($-0.14$ [$-0.22$, $-0.06$]). The redistribution therefore works through different ingredients in the two basins: weaker shear in the AS and weaker low-level vorticity in the BoB.

Sampling the environment along each storm's track links these large-scale changes to the intensification results of
Section~\ref{subsec:intensification}. BoB cyclonic storms in El Ni\~no seasons met about 40\% stronger shear between first reaching 34\,kt and reaching their LMI (14.9 versus 10.5\,m\,s$^{-1}$, $p=0.009$; Fig.~\ref{fig:mech}b), as expected for storms moving north into the part of
the Bay where El Ni\~no raises the shear (Fig.~\ref{fig:env}a). In all seasons together, BoB storms that reached VSCS met weaker along-track shear than those that did not (11.9 versus 15.0\,m\,s$^{-1}$, $p=0.033$). In the AS, along-track shear does not differ between the phases. Instead, El Ni~no-season AS storms tended to form about 3$^\circ$ further south (9.7$^\circ$N versus 12.7$^\circ$N, $p=0.06$; Fig.~\ref{fig:mech}c), in the warm, low-shear zone of Fig.~\ref{fig:env}. Starting further south would give them
longer tracks over water, which fits the duration-dominated ACE response in Fig.~\ref{fig:intensification}c. Ocean temperature adds to this in the AS. Monthly SST anomalies at the genesis position were higher for El Ni\~no-season
AS storms (+0.24\,\degC) than for La Ni\~na-season storms ($-0.08$\,\degC), and a one-standard-deviation warmer genesis SST raised the odds of an OND AS cyclonic storm reaching VSCS by a factor of 2.6 ($p=0.048$; $n=29$) the effect holds when the SST anomalies are linearly detrended (OR 3.0, $p=0.036$). BoB storms show no such SST effect (OR 0.85, $p=0.49$).

Monthly fields can account for only part of the storm-level ENSO effect. Adding along-track shear and humidity to the BoB storm model reduces the ENSO coefficient by about 14\%, and the 95\% bootstrap interval of this share includes zero. In the AS, along-track shear and humidity are unrelated to Ni~no-3.4 ($|\rho|<0.08$), so they carry none of the ENSO effect. The rest of the effect probably lies in synoptic conditions that monthly means cannot resolve.

The environmental fields also help to interpret the change over time described in Section~\ref{subsec:nonstationary}. The BoB vorticity response to ENSO was slightly weaker in the second half of the record ($\rho$ from $-0.54$ to $-0.42$; El Ni\~no minus La Ni\~na from $-0.37$ to $-0.24\times10^{-5}$,s$^{-1}$), in the same direction as the weaker BoB suppression since 2004, although the difference is not significant. The ENSO response of AS shear and humidity, on the other hand, did not grow ($\rho$ for shear from $-0.53$ to $-0.37$; for humidity from 0.36 to 0.26). The stronger AS TC response in recent El Ni\~no seasons is
therefore not the result of stronger ENSO forcing of the AS environment. It fits better with a warmer, and possibly moister, AS background, in which the same El Ni\~no-related drop in shear allows more storms to intensify. Over 1982--2025, OND 700\,hPa RH over the AS rose by 1.7\% per decade ($p=0.13$)
and OND SST by 0.16\,\degC\ per decade ($p<0.001$), while shear showed no trend ($p=0.72$). BoB SST warmed at the same rate (0.16\,\degC\,decade$^{-1}$, $p<0.001$), so SST warming by itself cannot explain why the increase in intense storms is confined to the AS.

\subsection{Seasonal predictability of the inter-basin partition}
\label{subsec:predict}

Because the ENSO phase is usually clear by August, the redistribution can be anticipated before the season begins. August Ni\~no-3.4 correlates with OND $\Delta \ACE$ at $\rho=0.63$ ($p\leq0.001$). A leave-one-out linear
hindcast reaches a cross-validated $r=0.41$ and an MSSS of 0.19, and it places 5 of the 11 top-quartile AS-dominated seasons in its own top quartile (HSS $=0.27$; Fig.~\ref{fig:stationarity}c). The same predictor has no skill
for total NIO OND ACE (MSSS $=-0.05$), as expected if ENSO redistributes energy rather than changing the total. The choice of skill measure matters here. $\Delta \ACE$ is negative in 80\% of seasons, so a forecast of ``BoB-dominated'' every year would be right 80\% of the time with no skill at all, and a sign hit rate says little. For seasonal outlooks, the practical message is that an ENSO forecast indicates which coast faces the greater post-monsoon intensification risk, but not how active the NIO season will be overall.

\section{Discussion}
\label{sec:discussion}

\subsection{What the results add}

\begin{table}[t]
\centering
\caption{Main results of this study compared with earlier work on the same questions. Earlier work is summarised only as far as it bears on each question.}
\label{tab:novelty}
\scriptsize
\begin{tabularx}{\textwidth}{@{}>{\raggedright\arraybackslash}p{2.9cm} >{\raggedright\arraybackslash}X >{\raggedright\arraybackslash}X@{}}
\toprule
\textbf{Topic} & \textbf{Earlier work} & \textbf{This study} \\
\midrule
ENSO and basin ACE & BoB suppressed in El Ni\~no \citep{girishkumar2012influences,bhardwaj2019tropical}; AS enhanced in OND \citep{almahri2025teleconnections}; opposite responses over 29 yr \citep{das2023tropical} & Tests redistribution against modulation; NIO total unchanged (GLM 0.99, $p=0.93$); 11 robustness tests \\
Life-cycle stage & Seasonal counts and ACE composited by ENSO phase; more intense BoB storms in La Ni\~na \citep{girishkumar2012influences,girishkumar2015influence,bhardwaj2019tropical}; weak RI--ENSO link in the Indian Ocean in a global analysis \citep{silvers2025global} & No detectable effect on formation; VSCS conversion OR 4.2 (AS) and 0.49 (BoB); separate duration and intensity pathways \\
Environmental pathway & Environments composited by ENSO phase for the BoB, where GPI terms point to vorticity and mid-level humidity \citep{girishkumar2015influence,bhardwaj2019tropical}, or for the AS alone \citep{almahri2025teleconnections} & Both basins in one framework: different GPI terms in each (AS shear, BoB vorticity); along-track shear and genesis-latitude evidence \\
ENSO vs IOD & Combined ENSO--IOD phase composites \citep{yuan2013north,mahala2015impacts,cheung2019comparison}; no ENSO or IOD contrast between active and inactive pre-monsoon seasons \citep{emmanuel2026suppression} & Post-monsoon IOD signal largely shared with ENSO (ENSO-independent DMI: $\rho=0.09$); no AMJ ENSO signal, but an ENSO-independent IOD link to AMJ AS intensification \\
Stationarity & ENSO--BoB link depends on PDO phase \citep{girishkumar2015influence}; results differ with the period covered \citep{tiwari2022assessing,thomas2022recent} & Significant change between 1982-2003 and 2004-2025  ($p=0.016$) robust to the split year; response moved from BoB suppression to AS enhancement \\
Trends and record length & Epochal AS increases \citep{deshpande2021changing,swapna2022increasing}; AS RI doubled after 2013 \citep{li2025sharp} & GLM trends with agency sensitivity; record-length dependence; no 15-yr window detects the AS trend \\
Genesis KDE & KDE climatology \citep{wahiduzzaman2017climatological,wahiduzzaman2020kernel} & Year-permutation and FDR test; epoch differences not field-significant \\
Prediction & Seasonal outlooks of NIO activity \citep{biswas2018principal,jaiswal2024seasonal} & Partition predictable (MSSS 0.19), total not; skill judged against an always-BoB reference \\
\bottomrule
\end{tabularx}
\end{table}

That ENSO affects NIO cyclones is not new, and Table~\ref{tab:novelty} sets our results beside the earlier work on each question. What the 44-year record adds is the \emph{form} of the effect. Taken over the whole period, post-monsoon cyclone energy in the NIO behaves, with respect to ENSO, almost as a fixed amount that ENSO shares out between the two sub-basins. El Ni\~no gives more of it to the AS and less to the BoB, and the basin total hardly changes. Studies that look at one sub-basin at a time see only half of this exchange, which may be why they can report a strong ENSO effect in one basin and a weak one in the other without either being wrong.

The second result is that ENSO acts after a storm has formed. The chance that a depression becomes a cyclonic storm shows no detectable dependence on ENSO in either basin. The odds that a cyclonic storm becomes very severe, on the other hand, rise about fourfold in the AS and halve in the BoB for each standard deviation of Ni\~no-3.4. This helps to explain why storm counts are only loosely related to ENSO while energy and the number of severe storms are closely related to it. It also means that seasonal counts are a poor measure of the ENSO signal in this basin. ACE and the number of
storms reaching VSCS carry most of the information.

The third result concerns time. The relationship we describe is not a fixed property of the basin. It was already known that the ENSO effect on BoB storms is stronger in some decades than in others, depending on the phase of the Pacific Decadal Oscillation \citep{girishkumar2015influence}. Our record shows a change of this kind between 1982--2003 and 2004--2025 and shows that it involves both basins. In 1982--2003 El Ni\~no mainly suppressed BoB storms and so lowered the NIO total. Since 2004 the AS gain in El Ni\~no seasons has grown large enough to balance the BoB loss. The rise of intense AS storms over the same decades \citep{murakami2017increasing,swapna2022increasing,li2025sharp} and the strengthening of the AS response to El Ni\~no may be two sides of one change: the AS has become more able to support intense storms, and El Ni\~no seasons are when that ability is used.

\subsection{Physical interpretation}
The environmental results suggest why the two basins respond in opposite ways, and why they do so through different parts of the storm life cycle. In El Ni\~no autumns, easterly anomalies develop at 850\,hPa over the equatorial Indian Ocean. Over the southern Bay these anomalies weaken the
cyclonic low-level circulation in which many BoB disturbances form, so fewer depressions develop \citep{roose2022enso}. The storms that do form meet stronger shear along their tracks than in La Ni\~na years, as they move north into the part of the Bay where El Ni\~no raises the shear, and they intensify less. The BoB is in fact slightly warmer in El Ni\~no autumns, so the suppression appears to come from the circulation rather than from the ocean \citep{bhardwaj2019tropical}.

Over the AS the same large-scale pattern has the opposite effect. Shear falls south of about 15$^\circ$N, the sea surface warms \citep{abhinav2025warming}, and the middle troposphere becomes somewhat moister. AS storms tend to form about 3$^\circ$ further south than in La Ni\~na years, start their lives in this favourable zone and
have a longer path over warm water before they reach land. The result is that AS storms last longer, rather than becoming much more numerous. The warmer AS surface in El Ni\~no autumns is consistent with the positive IOD events that often develop at the same time \citep{saji1999dipole}, but our data cannot separate the two in that season.

The two basins are therefore not mirror images. Different ingredients change in each (shear in the AS, low-level vorticity in the BoB), and those ingredients act on different aspects of storm development. A single basin-wide index such as the GPI would hide this difference. Our diagnostics also leave out potential intensity and ocean heat content, so the thermodynamic side of the picture rests on SST and humidity alone.

\subsection{Why studies disagree, and what this means for outlooks} Two results of this study may help to explain why published estimates for the NIO differ \citep{deshpande2021changing,tiwari2022assessing,thomas2022recent,abhinav2025warming}. The first is record length. Fifteen-year windows give AS energy trends of either sign, and none of them detects the long-term increase. Even the BoB, which has no trend over 1982--2025, shows significant decreases in windows of up to about three decades that start in the mid-to-late 1980s. A trend estimated from a short record can therefore
reflect where that record sits in the decadal variability as much as any long-term change. The second is the change in the ENSO relationship between the two halves of the record, which adds to earlier evidence that the ENSO effect on BoB storms varies between decades \citep{girishkumar2015influence}. In 1982--2003, El Ni\~no seasons had less total post-monsoon energy than La Ni\~na seasons; in 2004--2025 the total was no longer lower in El Niño seasons, and the AS response was stronger. Studies that cover different periods could therefore reach different conclusions about
ENSO even with the same data and methods. Both factors should be considered before differences between studies are attributed to data or method.

For seasonal outlooks the practical message is fairly simple. ENSO is usually well established by August \citep{mcphaden2006enso}, and its state then carries useful
information about \emph{which} sub-basin is more likely to produce intense post-monsoon storms, and hence which group of coastlines is more exposed: those around the AS (Oman, Pakistan, Gujarat and Maharashtra) or those around the BoB (Sri Lanka, Tamil Nadu, Andhra Pradesh, Odisha, Bangladesh
and Myanmar). Over the full record, it carries almost no information about how active the NIO season will be as a whole. Outlooks of basin-wide activity \citep{biswas2018principal,jaiswal2024seasonal} could therefore be complemented by an outlook of the inter-basin
partition. The skill is modest (cross-validated $r=0.41$), and because the BoB dominates in most years, any outlook of this kind should be judged against that reference and not against simple hit rates \citep{wilks2020statistical}.

\subsection{Limitations}
\label{subsec:limitations}
Several limitations should be kept in mind. AS samples are small: there are 29 OND and 24 AMJ AS cyclonic storms, and 15 OND seasons have no AS ACE at all. We have relied on rank statistics, resampling and storm-level models for this reason, and report confidence intervals throughout; the pre-monsoon IOD result in particular should be treated as a lead for
further work rather than a firm conclusion. Best-track intensities are not homogeneous. We harmonised IMD and JTWC winds and repeated the main tests with each agency separately, but some inhomogeneity, especially before satellite scatterometry, may remain. The comparison of the two halves of
the record rests on seven El Ni\~no and eight La Ni\~na seasons in each half. The change is significant and does not depend on the split year, but whether it is a response to warming or a product of internal variability can only be settled with longer records or large ensembles of model
simulations. The environmental analysis uses monthly fields, which cannot describe the conditions met by an individual storm on a given day, and it does not include potential intensity or upper-OHC; the latter is left for a follow-up study. Finally, all the relationships reported here are statistical associations and do not by themselves
establish cause.

\section{Conclusions}
\label{sec:conclusion}
This study used 44 years (1982--2025) of quality-controlled IBTrACS best tracks, with IMD and JTWC intensities harmonised, to examine how ENSO and the IOD affect TCs in the two sub-basins of the NIO. The number of cyclonic storms in the basin shows no trend since 1982, but the AS now produces many more intense storms: very severe cyclonic storms there increased by a factor of 1.58 per decade and rapid intensification events by 1.98 per decade, while no BoB metric shows a trend. None of the 30 possible 15-year windows detects the AS
increase, so trends in NIO cyclone activity need records of about three decades or more. The central result is that ENSO redistributes post-monsoon cyclone energy between the AS and the BoB rather than changing the basin total. For a one-standard-deviation increase in Ni\~no-3.4 (about 1\,\degC), AS ACE is multiplied by 1.89 (an increase of 89\%) and BoB ACE by 0.63 (a decrease of 37\%), while the NIO total shows no detectable change (factor 0.99 [0.74, 1.32]). ENSO acts after storms form: it changes the odds that a cyclonic storm becomes very severe, not the chance that a D becomes a storm. The two sub-basins respond through different ingredients, lower shear in the AS and weaker low-level vorticity in the BoB. In the post-monsoon season the IOD signal can hardly be separated from ENSO, whereas an ENSO-independent IOD influence appears in the pre-monsoon season. The relationship has also changed over time. In 1982--2003 El Ni\~no mainly suppressed BoB storms and lowered the NIO total; in 2004--2025 the AS gain balanced the BoB loss.

These results suggest that the two sub-basins are better treated as one coupled system, in seasonal outlooks and in coastal risk assessment. August Ni\~no-3.4 gives modest skill for the AS--BoB partition of post-monsoon energy but none for the basin total, so ENSO indicates which coastlines face
the greater risk in a given season rather than how active the season will be. Future work should include OHC and potential intensity, use daily environmental fields, and test with model ensembles whether the recent change in the ENSO relationship reflects warming or natural decadal
variability.

\section*{Data availability}
All data used in this study are publicly available. IBTrACS v04r01 best-track data \citep{knapp2010international} were obtained from NOAA NCEI (\url{https://www.ncei.noaa.gov/products/international-best-track-archive}). The monthly Ni\~no-3.4 and Dipole Mode Index series were obtained from NOAA PSL (\url{https://psl.noaa.gov/data/timeseries/month/}). ERSSTv5
\citep{huang2017extended} is available from NOAA PSL
(\url{https://psl.noaa.gov/data/gridded/data.noaa.ersst.v5.html}), and ERA5 \citep{hersbach2020era5} from the Copernicus Climate Change Service Climate Data Store (\url{https://cds.climate.copernicus.eu/}). The authors will share the processed storm list and the derived seasonal series with any reader who contacts the corresponding author.

\section*{Code availability}
The analysis was carried out in Python. The Python scripts used for preprocessing, statistical analysis and plotting can be obtained by writing to the corresponding author.

\section*{CRediT authorship contribution statement}
\textbf{Prashant Saini:} Conceptualisation, Methodology, Data filter and preprocessing, Formal analysis, Software, Visualisation, Writing -- original draft. \textbf{Rishikesh Yadav:} Conceptualisation, Methodology, Supervision, Validation, Writing -- review and editing. \textbf{Sarita Azad:} Conceptualisation, Supervision.

\section*{Funding}
Prashant Saini is supported by a research scholar fellowship from the Indian Institute of Technology Mandi. No further grant from any public, commercial or non-profit funding body supported this work.

\section*{Declaration of competing interest}
The authors have no financial or personal ties that could be seen as affecting the results or conclusions of this study.

\section*{Acknowledgements}
We thank the India Meteorological Department (RSMC New Delhi), the Joint Typhoon Warning Center and NOAA NCEI for the IBTrACS archive, NOAA PSL for the climate indices and ERSSTv5, and the Copernicus Climate Change Service for ERA5. We are grateful to Devanshu Rai and Kashish Chaudhary for their help with data handling and earlier analysis, and to G.~Sangeetha and Saurabh Kumar for in between useful discussions. We also thank the anonymous reviewers and the Associate Editor of our earlier submission, whose comments helped to improve the paper. AI-tools were used while preparing this work, for language editing and for checking sections of the analysis code. The authors examined and corrected all of its output and accept full responsibility for what the paper contains. The authors acknowledge
the School of Mathematical and Statistical Sciences, IIT Mandi, for computing facilities.

\appendix
\section{Supplementary figures}
\label{app:figures}
\renewcommand{\thefigure}{A.\arabic{figure}}
\renewcommand{\thetable}{A.\arabic{table}}
\setcounter{figure}{0}
\setcounter{table}{0}

\begin{figure}[!htbp]
\centering
\includegraphics[width=\textwidth]{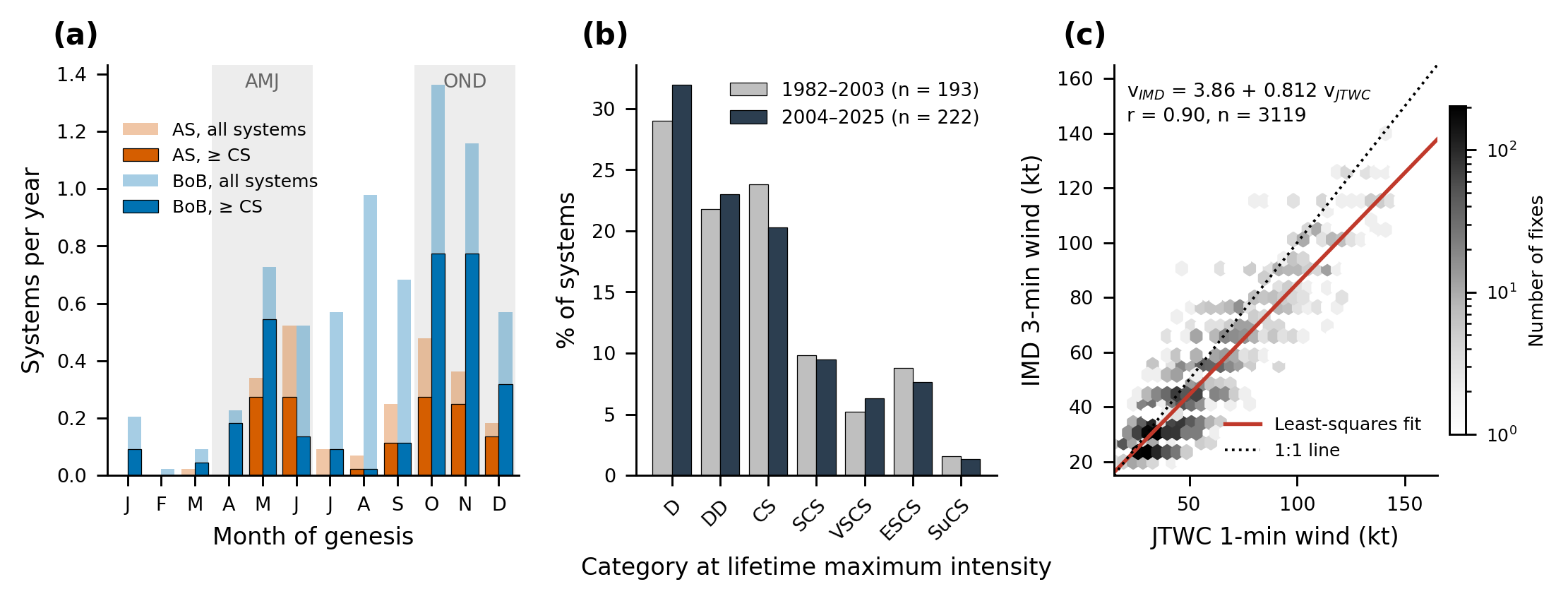}
\caption{(a) Monthly genesis climatology (systems per year, 1982--2025) in the AS and the BoB, for all systems (light bars) and for those that reached cyclonic-storm strength (dark bars); grey shading marks the AMJ and OND seasons. (b) Distribution of lifetime maximum intensity by category for
1982--2003 and 2004--2025 (percentage of systems; $n$ in the legend). (c) IMD versus JTWC winds at synoptic fixes where both are available and the IMD wind is at least 17\,kt, with the least-squares fit used for harmonisation (red line; Eq.~\ref{eq:conv}) and the 1:1 line (dotted).}
\label{fig:app_clim}
\end{figure}

\begin{figure}[!htbp]
\centering
\includegraphics[width=0.85\textwidth]{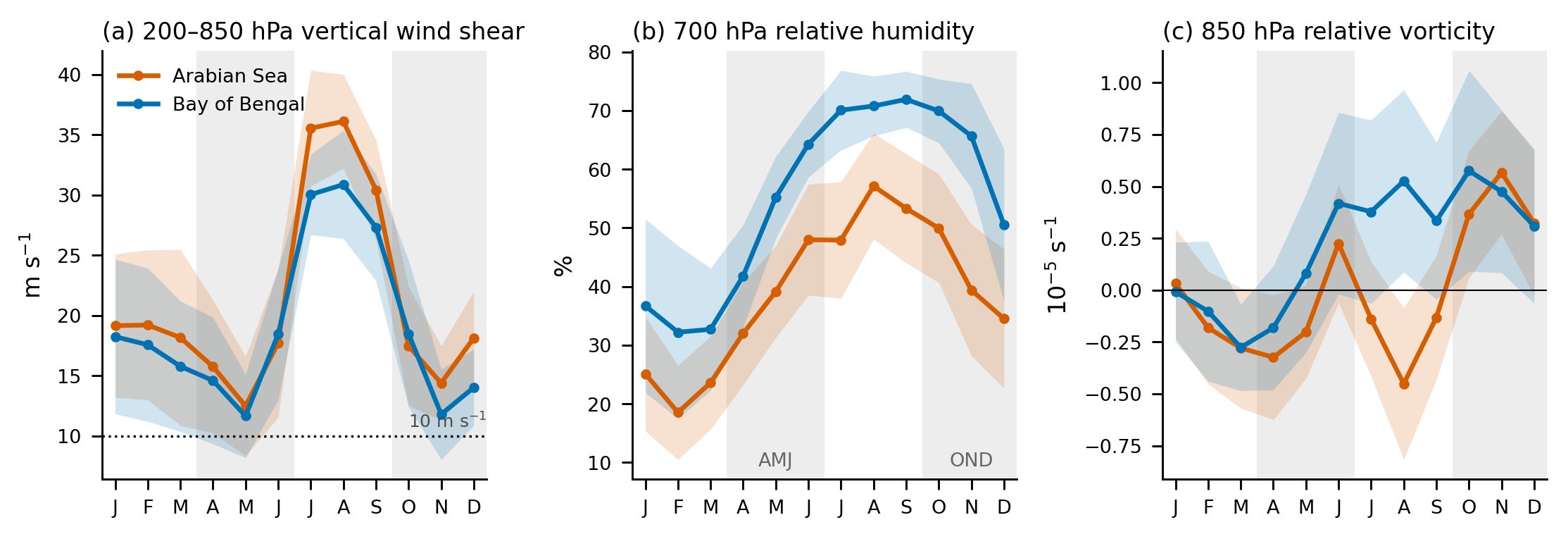}
\caption{Monthly climatology of (a) 200--850\,hPa VWS,
(b) 700\,hPa relative humidity and (c) 850\,hPa relative vorticity averaged over the AS and BoB boxes (ERA5, 1982--2025). Lines show the monthly mean and shading $\pm$1 standard deviation across years. Grey bands mark the AMJ
and OND cyclone seasons, and the dotted line in (a) marks a shear of 10\,m\,s$^{-1}$.}
\label{fig:app_env}
\end{figure}

This appendix gives two figures that support the data and methods. Fig.~\ref{fig:app_clim} shows the seasonal cycle of genesis, the distribution of storm intensity in the two halves of the record and the relation between IMD and JTWC winds used
to harmonise intensities. Fig.~\ref{fig:app_env} shows the seasonal cycle of the environmental variables averaged over the two analysis boxes.

\bibliographystyle{elsarticle-harv}
\bibliography{arXiv-paper/stat}

\end{document}